\documentclass[11pt,letterpaper]{article}

\usepackage[latin1]{inputenc}
\usepackage{amsmath}
\usepackage{amsfonts}
\usepackage{amssymb}
\usepackage{subfigure}
\usepackage[round, authoryear]{natbib}
\usepackage{mathrsfs}
\usepackage{graphicx}
\usepackage{xcolor}
\usepackage{url}
\usepackage[pagewise]{lineno}
\linenumbers

\newtheorem{theorem}{Theorem}

\newcommand{\argmax}{\operatornamewithlimits{argmax}}
\newcommand{\argmin}{\operatornamewithlimits{argmin}}

\newcommand{\bP}{\mathbf{P}}
\newcommand{\bQ}{\mathbf{Q}}

\newcommand{\bOmega}{\boldsymbol{\Omega}}
\newcommand{\bSigma}{\boldsymbol{\Sigma}}
\newcommand{\btheta}{\boldsymbol{\theta}}
\newcommand{\bbeta}{\boldsymbol{\beta}}

\title{The open-faced sandwich adjustment for MCMC using estimating functions}
\author{Benjamin A. Shaby}
\date{\today}

\begin{document}

\maketitle

\begin{abstract}
  The situation frequently arises where working with the likelihood function is
  problematic. This can happen for several reasons---perhaps the likelihood is
  prohibitively computationally expensive, perhaps it lacks some robustness
  property, or perhaps it is simply not known for the model under consideration.
  In these cases, it is often possible to specify alternative functions of the
  parameters and the data that can be maximized to obtain asymptotically normal
  estimates. However, these scenarios present obvious problems if one is
  interested in applying Bayesian techniques. Here we describe open-faced
  sandwich adjustment, a way to incorporate a wide class of non-likelihood
  objective functions within Bayesian-like models to obtain asymptotically valid
  parameter estimates and inference via MCMC. Two simulation examples show that
  the method provides accurate frequentist uncertainty estimates. The open-faced
  sandwich adjustment is applied to a Poisson spatio-temporal model to analyze
  an ornithology dataset from the citizen science initiative eBird.
\end{abstract}

\section{Introduction}
\label{sec:intro}

  For many models arising in various fields of statistical analysis, working
  with the likelihood function can be undesirable. This may be the case for
  several reasons---perhaps the likelihood is prohibitively expensive to
  compute, perhaps it presumes knowledge of a component of the model that one is
  unwilling to specify, or perhaps its form is not even known for a chosen
  probability model. Such scenarios present problems if one wishes to perform
  Bayesian analysis. Applying the Bayesian computational and inferential
  machinery, thereby enjoying benefits such as natural shrinkage, variance
  propagation, and the ability to incorporate complex hierarchical dependences,
  usually requires working directly with the likelihood function.
  
\begin{figure}[ht]
    \begin{center}
      \includegraphics[width=.75\linewidth, angle=0,
                       clip=true, trim=10 10 10 5]
                      {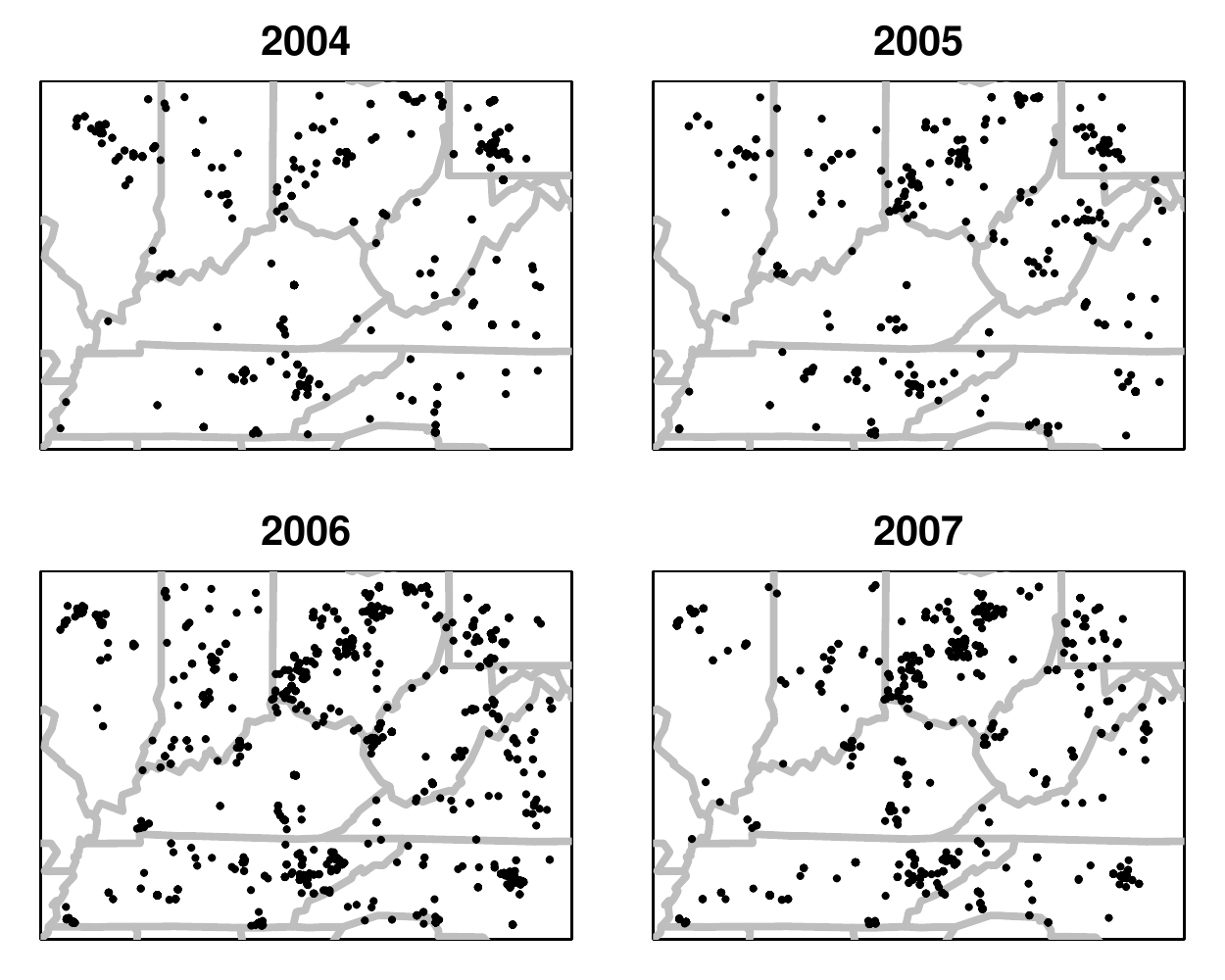}
    \end{center}
    \caption[]{Spatial locations of Northern Cardinal observations.}
    \label{fig:norcar-locations}
  \end{figure}

  To motivate the development, we briefly describe the analysis of bird
  sightings contained in Section \ref{sec:bird-counts}. The data consist of
  several thousand counts occurring irregularly in space and time (see Figure
  \ref{fig:norcar-locations}), along with several spatially-varying covariates
  carefully chosen by a group of ornithologists. A natural model for such data
  is a hierarchical Poisson regression with a random effect specified as a
  spatio-temporal Gaussian process with unknown covariance parameters. Here,
  whetever one's philosophical orientation, Bayesian methods are most practical
  to implement, and in addition provide sharing of information across space and
  time, as well as automatic uncertainty estimation of predictive abundance
  maps. Furthermore, obtaining an MCMC sample of the posterior distribution is
  desirable because inferences on the posterior correlation surface of the
  random effect, a nonlinear functional of random covariance parameters, is of
  independent interest to the ornithologists. However, the sheer size of the
  dataset makes MCMC under this model intractable, so a faster objective
  function is used in place of the high-dimensional Gaussian likelihood. The
  goal of the method presented here is to enable such a substitution while
  retaining a valid interpretation of the resultant MCMC sample.

  More generally, suppose that one specifies a model, either one-stage or
  hierarchical, and wants the advantages of being Bayesian, but the likelihood
  in some level of the hierarchy is problematic. Suppose, however, that one can
  write down some objective function $\ell_{M}(\boldsymbol{\theta}; \mathbf{y})$
  of the parameters and the data (possibly conditional on other parameters) that
  behaves similarly to the log likelihood. We will define what we mean by
  ``similarly'' in Section \ref{sec:framework}. Important examples of methods
  that employ such objective functions include generalized
  estimating equations \citep{hardin-2003a}, generalized method of moments
  \citep{hall-2005b}, and robust M-estimation \citep{huber-2009a}, as well as
  the two examples we will consider here, covariance tapering
  \citep{kaufman-2008b} and composite likelihoods \citep{lindsay-1988a}.
  
  The question we attempt to answer here is this: Can we insert
  $\ell_{M}(\boldsymbol{\theta}; \mathbf{y})$, in place of the likelihood, into
  an MCMC algorithm like Metropolis-Hastings and ``trick'' it into doing
  something useful? We claim that we can---that for many useful examples, simply
  swapping $\ell_{M}(\boldsymbol{\theta}; \mathbf{y})$ into a sampler results in
  a \emph{quasi}-posterior sample that can be rotated and scaled to yield
  desirable properties.
  
  The OFS adjustment relies on asymptotic theory that was formally developed in
  \citet{chernozhukov-2003a}, but which is quite intuitive. These authors were
  interested in using Metropolis-Hastings as an optimization algorithm for
  badly-behaved objective functions, not in using non-likelihood objective
  functions for performing Bayesian-like analysis, as we are here. Although their
  goals were entirely different, the theory contained therein is extremely
  useful for our purposes. 
 
  Previous attempts to incorporate non-likelihood objective functions into the
  Bayesian setting, to our knowledge, have been few. \citet{mcvean-2004a} use
  composite likelihoods within reversible jump MCMC, without any adjustment, to
  estimate population genetic parameters. Realizing that their sampler would
  result in invalid inferences, \citet{mcvean-2004a} turn to a parametric
  bootstrap to estimate sampling variability. \citet{smith-2009a} were
  interested in max-stable processes for spatial extreme value analysis. They
  also use composite likelihoods within MCMC without adjustment. The special
  case of using the generalized method of moments objective function
  \citep{hansen-1982a, hall-2005b} for generalized linear models within an MCMC
  sampler was explored by \citet{yin-2009a}. Tangentially related is
  \citet{tian-2007a}, who use MCMC to estimate the sampling distribution of
  $\hat{\boldsymbol{\theta}} = \argmax{\ell_{M}(\boldsymbol{\theta};
  \mathbf{y})}$. 
  
  \citet{ribatet-2012a} attempt to solve the same problem
  that we address here. Whereas we adjust quasi-posterior samples generated from
  MCMC post hoc, these authors propose an adjustment to the Metropolis
  likelihood ratio within the sampler itself. Their goal, like ours, is to
  achieve desirable frequentist coverage properties of credible intervals
  computed based on MCMC. Although their approach is quite general,
  \citet{ribatet-2012a} restrict their attention to using composite likelihoods
  for max-stable processes. The approach taken in the present article is closely
  related to that of \citet{ribatet-2012a}, but the OFS adjustment differs
  from their adjustment in its structure as well as its motivating asymptotic
  arguments.

  Both the motivating insights for the OFS adjustment and the criterion by which
  we evaluate it is essentially the idea of calibration \citep{draper-2006a}. In
  our interpretation, a well-calibrated method has the property that when used
  to construct credible intervals from many different datasets, those intervals
  ought to cover the true parameter at close to their nominal rates.
  Essentially, this says that well-calibrated credible intervals behave like
  confidence intervals. If we construct intervals with accurate coverage
  directly as the $\alpha/2$ and $(1-\alpha/2)$ empirical quantiles of an MCMC
  sample for different values of $\alpha$, we claim that in some way our
  uncertainty about a parameter is well-described by the sample. Evaluating an
  approximate Bayesian method by this criterion has intuitive practical appeal,
  and it has been endorsed in particular by objective Bayesians
  \citep[e.g.]{bayarri-2004a, berger-2001a}.
  
  This principle, along with some basic asymptotic observations, leads to the
  OFS adjustment. The asymptotic theory gives us the limiting normal
  distribution of quasi-Bayes point estimators. We take this distribution, in an
  informal sense, to be a summary of our uncertainty about $\btheta$, up to an
  asymptotic approximation. The asymptotic theory also gives us the limiting
  normal distribution of the quasi-posterior. Since these two limiting
  distributions are not, in general, the same, and since we would like the
  quasi-posterior to summarize our uncertainty about $\btheta$ in the sense of
  being well-calibrated, our strategy is to adjust samples from the
  quasi-posterior so that their limiting distribution matches that of the
  quasi-Bayesian point estimator.
  
  We note the temptation to ask how well the adjusted quasi-posterior
  distribution approximates the true posterior distribution, in cases when the
  true likelihood is available. However, this is the incorrect comparison to
  make. The true posterior distribution contains the information about $\btheta$
  obtained through the likelihood. When some other function $\ell_M$ is used in
  place of the likelihood, there is no reason to expect the information content
  to remain the same. We would like the adjusted quasi-posterior distribution to
  represent this loss of information, not hide it. In our simulation examples in
  Section \ref{sec:examples}, the frequentist accuracy of credible intervals
  based on adjusted quasi-posterior samples shows that the OFS adjustment
  accomplishes this task.

  Throughout, it will be assumed that expectations will be computed with respect
  to the true parameter $\boldsymbol{\theta}_0$. We define the square root of a
  symmetric positive definite matrix $\mathbf{A}$ to be $\mathbf{A}^{1/2}$ =
  $\mathbf{OD}^{1/2}\mathbf{O}'$, where $\mathbf{A=ODO'}$ with $\mathbf{O}$
  orthogonal and $\mathbf{D}$ diagonal. The square root of a matrix is not
  unique; here we compute $\mathbf{A}^{1/2}$ using the singular value
  decomposition, which is numerically stable and preserves key geometric
  attributes.

  We begin in Section \ref{sec:framework} by defining the quasi-Bayesian
  framework and reviewing the relevant asymptotic theory. In Section
  \ref{sec:ofs} we develop the OFS adjustment method, and we demonstrate how to
  apply it in two different statistical contexts in Section \ref{sec:examples}.
  In Section \ref{sec:data} we apply the OFS adjustment to analyze a
  dataset of Northern Cardinal sightings taken from the citizen science project
  eBird.  Section \ref{sec:discussion} concludes.

\section{The quasi-Bayesian framework}
\label{sec:framework}

  We begin by assuming that the parameter of interest $\boldsymbol{\theta}$
  lies in the interior of a compact convex space $\boldsymbol{\Theta}$.
  Suppose we are given $\mathbf{y}$, which consists of $n$ observations, from
  which we wish to estimate $\boldsymbol{\theta}$. Suppose further that we have
  at our disposal some objective function $\ell_{M}(\boldsymbol{\theta};
  \mathbf{y})$ from which it is possible to compute $\hat{\boldsymbol{\theta}}_M
  = \argmax_{\boldsymbol{\Theta}}{\ell_{M}(\boldsymbol{\theta}; \mathbf{y})}$.
  
  Following \citet{chernozhukov-2003a}, we define the
  \emph{quasi-posterior} distribution based on $n$ observations as
  \begin{equation}
    \pi_{M, n}(\boldsymbol{\theta}|\mathbf{y}_n) = 
      \frac{L_{M, n}(\boldsymbol{\theta}; \mathbf{y}_n)
            \pi(\boldsymbol{\theta})}
           {\int_\Theta 
             L_{M, n}(\boldsymbol{\theta}; \mathbf{y}_n)
             \pi(\boldsymbol{\theta})\,
             \mathrm{d}\boldsymbol{\theta}},
    \label{eqn:quasi-posterior}
  \end{equation}
  where $L_{M, n}(\boldsymbol{\theta}; \mathbf{y}_n) = \exp{\{\ell_{M,
  n}(\boldsymbol{\theta}; \mathbf{y}_n)\}}$, and $\pi(\boldsymbol{\theta})$ is a
  prior density on $\boldsymbol{\theta}$. We will assume, for convenience, that
  $\pi(\boldsymbol{\theta})$ is proper with support on $\boldsymbol{\Theta}$.
  The function $L_{M, n}$ is \emph{not} necessarily a density, and thus $\pi_{M,
  n}(\boldsymbol{\theta}|\mathbf{y}_n)$ is not a true posterior density in any
  probabilistic sense. We will assume, however, that $L_{M, n}$ is integrable,
  so as long as the prior $\pi(\boldsymbol{\theta})$ is proper, it easily
  follows that $\pi_{M, n}(\boldsymbol{\theta}|\mathbf{y}_n)$ will be a proper
  density.

  Equipped with notion of a quasi-posterior density, we can define 
  quasi-posterior risk as
  $R_n(\boldsymbol{\theta}) = \int_{\boldsymbol{\Theta}}
      \!\rho_n(\boldsymbol{\theta} - \boldsymbol{\theta}^*)
      \pi_{M,n}(\boldsymbol{\theta}^*|\mathbf{y}_n)
      \,\mbox{d}\boldsymbol{\theta}^*$,
  where $\rho_n(\mathbf{u})$ is some convex scalar loss function. For
  simplicity, we assume that $\rho_n(\mathbf{u})$ is symmetric, although this
  assumption may be dropped. Then for a given loss function, the quasi-Bayes
  estimator is naturally defined as $\hat{\boldsymbol{\theta}}_{\text{QB}}$
  $\,=\argmin_{\boldsymbol{\theta}\in\boldsymbol{\Theta}}
  R_n(\boldsymbol{\theta})$, the value of $\boldsymbol{\theta}$ that minimizes
  quasi-posterior risk.
    
  Our requirements on $\ell_{M,n}(\boldsymbol{\theta}; \mathbf{y}_n)$ are fairly
  minimal and are met by most objective functions in wide use in statistics.
  Technical assumptions are contained in \citet{chernozhukov-2003a}, but they
  are in general satisfied when $\hat{\boldsymbol{\theta}}_M$ is weakly
  consistent for $\boldsymbol{\theta}_0$ and asymptotically normal.

  Asymptotic normality of $\hat{\boldsymbol{\theta}}_M$ is of the form 
  \begin{equation}
    \label{eqn:asymp-theta-hat}
    \mathbf{J}_n^{1/2}
    (\hat{\boldsymbol{\theta}}_{M,n} - \boldsymbol{\theta}_0) 
    \xrightarrow{\mathscr{D}} N(\mathbf{0, I}),
  \end{equation}
  where
  \begin{align}
    \label{eqn:important-matrices}
    \mathbf{J}_n & =
    \mathbf{Q}_n \mathbf{P}^{-1}_n \mathbf{Q}_n
    \nonumber \\
    \mathbf{P}_n & =
    \mbox{E}_0[\nabla_0\ell_{M,n}
               \nabla_0\ell_{M,n}'] 
    \nonumber \\
    \mathbf{Q}_n & =
    -\text{E}_0[\mathcal{H}_0\ell_{M, n}].
  \end{align}
  The notation $\nabla_0 f$ refers to the gradient of the
  function $f$ evaluated at the true parameter
  $\boldsymbol{\theta}_0$, and $\mathcal{H}_0 f$ refers to the Hessian
  of $f$ evaluated at $\boldsymbol{\theta}_0$.  These matrices have
  been defined in terms of partial derivatives, but in general,
  $\ell_{M, n}$ does not have to be differentiable or even continuous
  for the theory to apply.  In this case, small adjustments of the
  definitions of $\mathbf{P}_n$ and $\mathbf{Q}_n$ are necessary.

  The sandwich matrix $\mathbf{J}^{-1}_n$ is familiar from generalized
  estimating equations, quasi-likelihood, and other areas, and is referred to by
  various names, including the \emph{Godambe information criterion} and the
  \emph{robust information criterion} \citep[e.g.][]{durbin-1960a,
  bhapkar-1972a, morton-1981a, ferreira-1982a, godambe-1987a, heyde-1997a}. We
  note that in the special case when $\ell_{M,n}(\boldsymbol{\theta};
  \mathbf{y})$ is the true likelihood, $\mathbf{Q}_n \equiv \mathbf{J}_n$, the
  Fisher information. We will hereafter assume that this is not the case.

\subsection{Review of relevant asymptotic theory}
\label{sec:asymptotics}


  \citet{chernozhukov-2003a} elucidates the asymptotic behavior of
  $\pi_{M,n}(\boldsymbol{\theta}|\mathbf{y}_n)$, which motivates the open-face
  sandwich adjustment. These results are direct analogues of well-known
  asymptotic properties of true posterior distributions. Their Theorem 2, which
  we re-state below, states that the asymptotic distribution of the quasi-Bayes
  estimator $\hat{\boldsymbol{\theta}}_{\text{QB},n}$ is the same as that of the
  extremum estimator $\hat{\boldsymbol{\theta}}_{M,n}$.
  
  \begin{theorem}
    \label{thm:quasi-Bayes-normal}
    Assuming sufficient regularity of $\ell_{M,n}(\boldsymbol{\theta};
    \mathbf{y}_n)$,
    \begin{displaymath}
      \mathbf{J}_n^{1/2}(\hat{\boldsymbol{\theta}}_{\text{QB}} -
                       \boldsymbol{\theta}_0)
      \xrightarrow{\mathscr{D}} N(\mathbf{0, I}).
    \end{displaymath}
  \end{theorem}  
  Theorem \ref{thm:quasi-Bayes-normal} above is the quasi-posterior extension of the
  well-known result that, under fairly general conditions, Bayesian point
  estimates have the same asymptotic distribution as maximum likelihood
  estimates.
  
  Theorem 1 of \citet{chernozhukov-2003a}, which we re-state here in a slightly
  different form, is a kind of quasi-Bayesian consistency result, showing that
  quasi-posterior mass accumulates at the true parameter
  $\boldsymbol{\theta}_0$.

  \begin{theorem}
    \label{thm:quasi-posterior-convergence}
    Under the same conditions as Theorem \ref{thm:quasi-Bayes-normal}
    \begin{displaymath}
      \| \pi_{M,n}(\boldsymbol{\theta}|\mathbf{y}_n) - 
         \pi_{M,\infty}(\boldsymbol{\theta}|\mathbf{y}_n) \|_{\text{TV}}
      \xrightarrow{\mathscr{P}} 0,
    \end{displaymath}
    where $\| \cdot \|_\text{TV}$ indicates the total variation norm, and
    $\pi_{M,\infty}(\boldsymbol{\theta}|\mathbf{y}_n)$ is a normal density with
    random mean $\boldsymbol{\theta}_0 +
    \mathbf{Q}_n^{-1}\nabla\boldsymbol{\ell}_{M,n}(\boldsymbol{\theta}_0)$ and
    covariance matrix $\mathbf{Q}_n^{-1}$.
  \end{theorem}
  Theorem \ref{thm:quasi-posterior-convergence} may be arrived at informally via
  a simple Taylor series argument. It is therefore intuitive that the
  quasi-posterior converges to limiting normal distribution whose covariance
  matrix is defined by the second derivatives of $\ell_\text{M}$.
 
  The key observation is that the limiting quasi-posterior distribution has a
  \emph{different} covariance matrix than the asymptotic sampling distribution
  of the quasi-Bayes point estimate. The consequence is that the usual Bayesian
  method of constructing credible intervals based on quantiles of the
  quasi-posterior sample will, viewed as confidence intervals, not have their
  nominal frequentist coverage probabilities. Fortunately, thanks to
  \citet{chernozhukov-2003a}, we know what those two asymptotic covariance
  matrices look like, which suggests a way to ``fix''
  $\pi_{M,n}(\boldsymbol{\theta}|\mathbf{y}_n)$.

\section{The open-faced sandwich adjustment}
\label{sec:ofs}

  Let us assume that we have a sample of draws from
  $\pi_{M,n}(\boldsymbol{\theta}|\mathbf{y}_n)$, generated by replacing the
  likelihood with $\ell_M(\boldsymbol{\theta}; \mathbf{y})$ in some MCMC sampler
  such as Metropolis-Hastings. Our aim here is to adjust the quasi-posterior
  draws such that the adjusted sample realistically reflects how the data
  informs our uncertainty about the parameter of interest $\boldsymbol{\theta}$
  through the function $\ell_M(\boldsymbol{\theta}; \mathbf{y})$. Were that the
  case, the usual credible intervals constructed from empirical quantiles of the
  adjusted sample would have close to nominal coverage. We will
  accomplish this by constructing a matrix $\boldsymbol{\Omega}_n$ that, when
  applied to the (centered) quasi-posterior sample, will rotate and scale the
  points in an appropriate way.
  
  We have observed that whereas the asymptotic covariance matrix of
  $\hat{\boldsymbol{\theta}}_{M,n}$ is the sandwich matrix $\mathbf{J}_n^{-1}$,
  the asymptotic covariance matrix of the quasi-posterior distribution
  is a single ``slice of bread'' $\mathbf{Q}_n^{-1}$. What we want to do then is
  complete the sandwich by joining the slice of bread $\mathbf{Q}_n^{-1}$ to the
  open-faced sandwich $\mathbf{P}_n\mathbf{Q}_n^{-1}$ to get
  $\mathbf{J}_n^{-1}$.

  We define $\boldsymbol{\Omega}_n =
  \mathbf{Q}_n^{-1}\mathbf{P}_n^{1/2}\mathbf{Q}_n^{1/2}$, the open-faced
  sandwich adjustment matrix. One can easily check that if $\mathbf{Z}_n
  \sim N(\mathbf{0}, \mathbf{Q}_n^{-1})$, then
  $\boldsymbol{\Omega}_n\mathbf{Z}_n \sim N(\mathbf{0}, \mathbf{J}_n^{-1})$. The
  idea then is to take samples from
  $\pi_{M}(\boldsymbol{\theta}|\mathbf{y})$ obtained via MCMC and
  pre-multiply them (after centering) by an estimator
  $\hat{\boldsymbol{\Omega}}$ of $\boldsymbol{\Omega}$ to ``correct'' the
  quasi-posterior sample. That is, if $\boldsymbol{\theta}^{(1)}, \ldots,
  \boldsymbol{\theta}^{(J)}$ is a sample from
  $\pi_{M}(\boldsymbol{\theta}|\mathbf{y})$, then for each $j=1, \ldots, J$,
  \begin{equation}
    \label{eqn:ofs-adj}
    \boldsymbol{\theta}_\text{OFS}^{(j)} = 
    \hat{\boldsymbol{\theta}}_\text{QB} +
    \hat{\boldsymbol{\Omega}}(\boldsymbol{\theta}^{(j)} -
    \hat{\boldsymbol{\theta}}_\text{QB})
  \end{equation}
  is the open-face sandwich adjusted sample. It is clear that a consistent
  estimator of $\bOmega$ will generate credible intervals that are consistent
  $(1-\alpha)$ confidence intervals.

\subsection{Estimating $\boldsymbol{\Omega}$}
\label{sec:estimating-Omega}

  The OFS adjustment \eqref{eqn:ofs-adj} requires an estimate of the matrix
  $\boldsymbol{\Omega}$, which in turn requires estimates of $\mathbf{P}$ and
  $\mathbf{Q}$. Because the OFS adjustment occurs post-hoc, it is possible to
  leverage the existing MCMC sample to compute $\hat{\boldsymbol{\Omega}}$.
  There are many possible approaches to this task, and here we offer some
  suggestions, which we summarize in Table \ref{tab:omega-estimators}.

  While is $\mathbf{P}$ is notoriously difficult to estimate well
  \citep[see][for some examples]{kauermann-2001a}, Theorem
  \ref{thm:quasi-Bayes-normal} immediately suggests a way to estimate
  $\mathbf{Q}$ directly from the MCMC sample with almost no additional
  computational cost. Specifically, noting that the quasi-posterior density
  converges to a normal with covariance matrix $\mathbf{Q}^{-1}$, a natural
  estimate $\hat{\mathbf{Q}}^{-1}_\text{I}$ is just the sample covariance matrix
  of the MCMC sample. Another possibility that requires almost no additional
  computation is to retain the results of the evaluations of $\ell_{M}$ at each
  iteration of the sampler and use them to numerically estimate the Hessian
  matrix at $\hat{\boldsymbol{\theta}}_{\text{QB}}$. This Hessian approximation
  will generally be a good estimator $\hat{\mathbf{Q}}_\text{II}$ of
  $\mathbf{Q}$.
  
  These estimators of $\mathbf{Q}$ are not only simple to compute, but they
  arise as direct results of MCMC output, requiring no additional analytical
  derivations based on $\ell_{M}$. They are, in this sense, ``model-blind.''
  Unfortunately, we are unaware of any such ``model-blind'' estimators of
  $\mathbf{P}$. The simplest solution, in the case where we can write an
  expression for $\nabla\ell_{M}(\boldsymbol{\theta}; \mathbf{y})$ and the data
  $\mathbf{y}$ consists of $n$ independent replicates, is to compute a basic
  moment estimator
  \begin{equation}
    \label{eqn:P-hat-moment}
    \hat{\mathbf{P}}_\text{I} = \frac{1}{n} \sum_{i=1}^n 
      \nabla\ell_{M}(\hat{\boldsymbol{\theta}}_{\text{QB}}; \mathbf{y}_i)
      \nabla\ell_{M}(\hat{\boldsymbol{\theta}}_{\text{QB}}; \mathbf{y}_i)',
  \end{equation}
  which is consistent as $n \rightarrow \infty$ under standard regularity
  conditions. We use equation \eqref{eqn:P-hat-moment} in Section
  and \ref{sec:composite}, where we have replication. However, because
  $\nabla\ell_{M}(\hat{\boldsymbol{\theta}}_{\text{QB}}; \mathbf{y})$ converges
  to zero, when we only observe a single realization of a stochastic process, as
  in Section \ref{sec:tapering}, equation \eqref{eqn:P-hat-moment} fails to
  provide a viable estimator. In this latter example, analytical expressions for
  $\mathbf{P}(\boldsymbol{\theta})$ are available. Plugging
  $\hat{\boldsymbol{\theta}}_{\text{QB}}$ into the analytical asymptotic
  expression gives an estimator $\hat{\mathbf{P}}_\text{II}$. If a corresponding
  analytical expression exists for $\mathbf{Q}$, we call the corresponding
  plug-in estimator $\hat{\mathbf{Q}}_\text{III}$
  
  When an expression for $\mathbf{P}(\boldsymbol{\theta})$ is unavailable, but
  when it is possible to simulate the process that generated $\mathbf{y}$, the
  parametric bootstrap is an attractive option. Let $\mathbf{y}_1, \ldots,
  \mathbf{y}_K$ be $K$ independent realizations of the stochastic process
  generated under $\hat{\boldsymbol{\theta}}_{\text{QB}}$. Then   
  \begin{equation}
    \label{eqn:P-hat-bootstrap}
    \hat{\mathbf{P}}_\text{boot} = \frac{1}{K} \sum_{k=1}^K 
      \nabla\ell_{M}(\hat{\boldsymbol{\theta}}_{\text{QB}}; \mathbf{y}_k)
      \nabla\ell_{M}(\hat{\boldsymbol{\theta}}_{\text{QB}}; \mathbf{y}_k)' 
  \end{equation}
  is the parametric bootstrap estimator of $\mathbf{P}$ (an analogous estimator
  could, of course, be used for $\mathbf{Q}$). A nice feature of
  \eqref{eqn:P-hat-moment} and \eqref{eqn:P-hat-bootstrap} is that, at the
  expense of (perhaps considerable) computational effort, one could substitute
  finite-difference approximations to the required gradients to obtain reasonable
  estimators, even in the absence of available closed-form expressions for
  $\nabla\ell_{M}(\boldsymbol{\theta}; \mathbf{y})$.
  
  \begin{table}[ht]
	  \centering
	  \begin{tabular}{ll}
	    Estimator & Description \\
	    \hline
		  $\hat{\bQ}_{\text{I}}^{-1}$ & sample covariance of MCMC sample \\
		  $\hat{\bQ}_{\text{II}}$ & Hessian of $\ell_M(\hat{\btheta}_\text{QB})$ \\
		  $\hat{\bQ}_{\text{III}}$ & plug $\hat{\btheta}_\text{QB}$ into asymptotic
		  formula\\
		  $\hat{\bP}_{\text{I}}$ & moment estimator based on score vector \\
		  $\hat{\bP}_{\text{II}}$ & plug $\hat{\btheta}_\text{QB}$ into asymptotic
		  formula\\
		  $\hat{\bP}_{\text{boot}}$ & parametric bootstrap \\
	  \end{tabular}
	  \caption{Summary of estimators of sandwich components.}
	  \label{tab:omega-estimators}
  \end{table}

\subsection{The curvature adjustment}
\label{sec:curvadj}

  We now describe the curvature-adjusted sampler of \citet{ribatet-2012a}. This
  sampler was presented as a way to include composite likelihoods in
  Bayesian-like models but in fact has far wider generality. Composite
  likelihoods \citep{lindsay-1988a} are functions of $\btheta$ and $\mathbf{y}$
  constructed as the product of joint densities of subsets of the data. In
  effect, composite likelihoods treat these subsets as though they were
  independent. Under fairly general regularity conditions, the asymptotic
  distribution of maximum composite likelihood estimators have sandwich form
  \eqref{eqn:asymp-theta-hat} \citep{lindsay-1988a}. Although
  \citet{ribatet-2012a} consider only composite likelihoods, their argument
  holds equally well for any function $\ell_M(\btheta; y)$ with sandwich
  asymptotics.
  
  The curvature-adjusted sampler begins by computing the extremum estimator
  $\hat{\btheta}_M$ and $\hat{\bOmega}(\hat{\btheta}_{M})$ as a preliminary
  step. It works by modifying the Metropolis-Hastings algorithm using a
  transformation of the form \eqref{eqn:ofs-adj} such that the acceptance ratio
  has a desirable asymptotic distribution. Specifically, at each iteration $j=1,
  \ldots, J$, the algorithm proposes a new value $\btheta ^*$ from some density
  $q(\cdot|\btheta^{(j)})$ and compares it to the current state $\btheta^{(j)}$
  to evaluate whether to accept or reject the proposal. The difference between
  the curvature-adjusted sampler and the traditional Metropolis-Hastings sampler
  is that this comparison now takes place after $\btheta ^*$ and $\btheta^{(j)}$
  are scaled and rotated. That is, $\btheta^*$ is accepted with probability
  \begin{equation}
    \label{eqn:curv-ratio}
    \min \bigg\{1, \frac{L_M(\btheta^*_{CA}; \mathbf{y})
                        \pi(\btheta^*)q(\btheta^{(j)}|\btheta^*)}
                       {L_M(\btheta^{(j)}_{CA}; \mathbf{y})
                        \pi(\btheta^{(j)})q(\btheta^*|\btheta^{(j)})}
         \bigg\}
  \end{equation}
  where $\btheta^*_{CA} = \hat{\btheta}_M +
  \hat{\bOmega}(\hat{\btheta}_{M})(\btheta^* - \hat{\btheta}_M)$, and
  analogously for $\btheta^{(j)}_{CA}$. \citet{ribatet-2012a} use a result from
  \citet{kent-1982a} to show that the ratio in \eqref{eqn:curv-ratio} has the
  same asymptotic distribution as that of the true likelihood ratio, and argue
  that the resultant sample has an asymptotic stationary distribution that is
  normal with covariance $\mathbf{J}^{-1}$, as desired. Note that unlike the OFS
  adjustment, the curvature-adjusted sampler requires outside initial estimates
  of $\btheta$ and $\bOmega$ because the adjustment occurs within the sampling
  algorithm.

\subsection{OFS within a Gibbs sampler}
\label{sec:ofs-gibbs}

  With some care, the OFS adjustment may be applied in the Gibbs sampler
  setting. Suppose we divide $\boldsymbol{\theta}$ into $B$ blocks such that
  $\boldsymbol{\theta}_1, \ldots, \boldsymbol{\theta}_B$ forms
  a partition of $\boldsymbol{\theta}$, and we wish to draw from the quasi-full
  conditional distribution with density $f(\boldsymbol{\theta}_{i} |
  \boldsymbol{\theta}_{-i}, \mathbf{y}) \propto L_{M}(\mathbf{y} |
  \boldsymbol{\theta}) f(\boldsymbol{\theta}_{i})$, where
  $\boldsymbol{\theta}_{-i}$ refers to the elements of $\boldsymbol{\theta}$ not
  contained in $\boldsymbol{\theta}_{i}$. Then the adjustment matrix
  $\boldsymbol{\Omega}_{\boldsymbol{\theta}_{i}|\boldsymbol{\theta}_{-i}}$ is
  defined as before, only now it applies only to $\btheta_i$ and is conditional
  on $\btheta_{-i}$. If all full conditional densities
  $f(\boldsymbol{\theta}_{i} | \boldsymbol{\theta}_{-i}, \mathbf{y})$, $i=1,
  \ldots, B$, are quasi-full conditional densities in the sense that they are
  proportional to a product of $\ell_{M}(\mathbf{y} | \boldsymbol{\theta})$ and
  another density, the Gibbs sampler may be run by successively drawing from
  $f(\boldsymbol{\theta}_{i} | \boldsymbol{\theta}_{-i}, \mathbf{y}), i=1,
  \ldots B$, and the OFS adjustment may proceed post hoc as before. This can be
  seen by viewing the Gibbs sampler as a special case of Metropolis-Hastings
  \citep[see][Section 7.1.4]{robert-2004a}. 
  
  Now suppose that at iteration $j$ of a Gibbs sampler we have drawn
  $\boldsymbol{\theta}_{i}^{(j)}$ from the quasi-full conditional
  $f(\boldsymbol{\theta}_{i} | \boldsymbol{\theta}_{-i}^{(j)}, \mathbf{y})$.
  Suppose further that $f(\boldsymbol{\theta}_{i+1} |
  \boldsymbol{\theta}_{-(i+1)}^{(j)}, \mathbf{y})$ is \emph{not} a function of
  $L_{M}$, as will be the case for many parameters in hierarchical models that
  contain $L_{M}$. Since $f(\boldsymbol{\theta}_{i+1} |
  \boldsymbol{\theta}_{-(i+1)}^{(j)}, \mathbf{y})$ is a true full conditional
  density and not a quasi-full conditional density, there is no OFS adjustment
  to make. But clearly $f(\boldsymbol{\theta}_{i+1} |
  \boldsymbol{\theta}_{-(i+1)}^{(j)}, \mathbf{y})$ depends on
  $\boldsymbol{\theta}_{i}^{(j)}$, and as a result, plugging in un-adjusted
  samples of $\boldsymbol{\theta}_{i}^{(j)}$ will not result in the desired
  stationary distribution. It is clear then that to achieve proper variance
  propagation through the model, we must adjust $\boldsymbol{\theta}_{i}^{(j)}$
  \emph{before} plugging it into $f(\boldsymbol{\theta}_{i+1} |
  \boldsymbol{\theta}_{-(i+1)}^{(j)}, \mathbf{y})$. Therefore, the OFS
  adjustment within the Gibbs sampler may not be applied post hoc, but rather
  must occur within the sampling algorithm.

  Embedding OFS adjustments within Gibbs samplers requires careful consideration
  of the conditional OFS matrices $\bOmega_{\btheta_{i} | \btheta_{-i}}$, $i=1,
  \ldots, B$, the adjustment matrices associated with the quasi-full conditional
  distributions $f(\btheta_i | \btheta_{-i})$, $i=1, \ldots, B$. Because it is
  defined conditionally on $\btheta_{-i}$, ideally each $\bOmega_{\btheta_{i} |
  \btheta_{-i}}$ should be re-estimated at each iteration $j$ based on the
  current value $\boldsymbol{\theta}_{-i}^{(j)}$. We refer to
  $\hat{\bOmega}^{(j)}_{\btheta_i | \btheta_{-i}}$ as the conditional OFS
  adjustment matrix for $\boldsymbol{\theta}_{i}^{(j)}$ at iteration $j$.
  Implementing the OFS Gibbs sampler using these conditional OFS adjustments
  requires estimation of $\hat{\bOmega}^{(j)}_{\btheta_i | \btheta_{-i}}$,
  by one of the techniques described in Section \ref{sec:estimating-Omega},
  for each block of parameters $i=1, \ldots, B$ with corresponding quasi-full
  conditional depending on $L_M$, and for each Gibbs iteration $j$ in $1,
  \ldots, J$. Furthermore, each computation of $\hat{\bOmega}^{(j)}_{\btheta_i |
  \btheta_{-i}}$ requires an estimate of $\boldsymbol{\theta}_{i}^{(j)}$ as input,
  necessitating some sort of optimization or MCMC, again nested within each
  block $i=1, \ldots, B$ and each iteration $j$ in $1, \ldots, J$. This is an
  enormous computational burden.
  
  Instead, we make the simplifying assumption that $\bOmega_{\btheta_{i} |
  \btheta_{-i}}^{(j)}$ does not change much from iteration to iteration. We
  instead use a constant (with respect to $j$) estimate
  $\hat{\bOmega}_{\btheta_{i} | \hat{\btheta}_{-i}}$, where $\hat{\btheta}_{-i}$ is
  fixed at its marginal quasi-Bayes estimate. We refer to
  $\hat{\bOmega}_{\btheta_{i} | \hat{\btheta}_{-i}}$ as the marginal OFS
  adjustment matrix for $\boldsymbol{\theta}_{i}^{(j)}$.
    
  Despite this simplification, the OFS-adjusted Gibbs sampler still requires
  additional work relative to an an-adjusted sampler because the algorithm
  requires $\hat{\boldsymbol{\theta}}_\text{QB}$ and
  $\hat{\boldsymbol{\Omega}}_{\boldsymbol{\theta}_{i} |
  \hat{\boldsymbol{\theta}}_{-i}}$, $i=1, \dots, B$, as input. In practice,
  then, we run the sampler twice. The first time, recalling that Theorem
  \ref{thm:quasi-posterior-convergence} says the un-adjusted quasi-posterior
  concentrates its mass at $\boldsymbol{\theta}_0$, we make no OFS adjustments,
  and use the generated sample to produce $\hat{\boldsymbol{\theta}}_\text{QB}$.
  We next use $\hat{\boldsymbol{\theta}}_\text{QB}$ to produce
  $\hat{\boldsymbol{\Omega}}_{\boldsymbol{\theta}_{i} |
  \hat{\boldsymbol{\theta}}_{-i}}$, $i=1, \dots, B$, using one of the methods
  described in Section \ref{sec:estimating-Omega}. Finally, the Gibbs sampler is
  re-run, this time with the transformation defined by \eqref{eqn:ofs-adj}
  applied for each block $i$ and each iteration $j$, $i=1, \ldots, B$, $j=1,
  \ldots, J$. Thus, the computational burden required to use marginal OFS
  adjustments is approximately twice that of the un-adjusted Gibbs sampler. In
  contrast, the additional computational burden required to use conditional OFS
  adjustments may range from several fold to several thousand fold, depending on
  the method used to estimate $\bOmega^{(j)}_{\btheta_i | \btheta_{-i}}$.
  
  We have explored (informally) the effects of using the much more
  computationally efficient marginal adjustment instead of the conditional
  adjustment and found only very minor differences in the resultant adjusted
  quasi-posteriors. (See Section \ref{sec:tapering} for an example.) The issue
  of conditional vs. marginal adjustments also appears in \citet{ribatet-2012a}.
  They refer to using constant (in $j$) adjustment matrices as an ``overall''
  Gibbs sampler and using conditional adjustment matrices as an ``adaptive''
  Gibbs sampler. Corroborating our findings, \citet{ribatet-2012a}, in a more
  systematic study using a very simple model, found very little
  difference between their overall and adaptive curvature-adjusted
  quasi-posteriors.

\section{Examples}
\label{sec:examples}

  We now describe two examples of non-likelihood objective functions that have
  appeared in the literature. In each example, working with the likelihood is
  problematic for a different reason, and each fits into the OFS framework. In
  the first example, we apply covariance tapering \citep{furrer-2006a,
  kaufman-2008b, shaby-2012a} to large spatial datasets. Here the likelihood
  requires the numerical inversion of a very large matrix. For large datasets,
  this inversion becomes prohibitively computationally-expensive, so the
  likelihood is replaced with its tapered version, which leverages sparse-matrix
  algorithms to speed up computations.  In the second example,
  composite likelihoods for spatial max-stable processes
  \citep{smith-1990a,padoan-2010a}, a probability model is assumed, but the
  likelihood consists of a combinatorial explosion of terms, and is therefore
  completely intractable for all but trivial situations. Hence, in this example,
  the likelihood function is simply not known. These examples may be considered
  toy models in that one could easily maximize their associated objective
  functions and compute sandwich matrices to obtain point estimates and
  asymptotic confidence intervals. We use these examples simply to illustrate
  the effectiveness of the OFS framework.
  
  For each example in this section, we conduct a simulation study to investigate
  how well the OFS adjustment performs by measuring how often nominal
  $(1-\alpha)$ credible intervals cover $\boldsymbol{\theta}_0$. To do this,
  we draw datasets $\mathbf{y}_k$, $k=1, \ldots, 1000$, from the model
  determined by some fixed $\boldsymbol{\theta}_0$. We then run a random walk
  Metropolis algorithm, with $\ell_M(\boldsymbol{\theta}; \mathbf{y}_k)$
  inserted in place of a likelihood, on each of the 1000 datasets. Next, we use
  each set of MCMC samples to compute estimates
  $\hat{\boldsymbol{\theta}}_{\text{QB},k}$ and $\hat{\boldsymbol{\Omega}}_k$
  using different estimators as discussed in Section \ref{sec:estimating-Omega}.
  Finally, we use each $\hat{\boldsymbol{\theta}}_{\text{QB},k}$ and
  $\hat{\boldsymbol{\Omega}}_k$ to adjust their corresponding batch of MCMC
  output, and record the resultant equi-tailed $(1-\alpha)$ credible interval
  for many values of $\alpha$. In addition, we run the curvature-adjusted
  sampler of \citet{ribatet-2012a} for comparison. For each example, empirical
  coverage rates are plotted against nominal coverage probabilities.

\subsection{Tapered likelihood for spatial Gaussian processes}
\label{sec:tapering}

  The most common structure for modeling spatial association among observations
  is the Gaussian process \citep{cressie-1991a, stein-1999a}. In addition to
  modeling Gaussian responses, the Gaussian process has been used extensively in
  hierarchical models to induce spatial correlation for a wide variety of
  response types \citep{banerjee-2004a}.
  
  Here we assume that $Y(\mathbf{s}) \sim \text{GP}(0, C(\boldsymbol{\theta});
  \mathbf{s})$, a mean-zero Gaussian process whose second-order stationary
  covariance is given by a parametric family of functions $C$ indexed by
  $\boldsymbol{\theta}$, depending on locations $\mathbf{s}$ in some spatial
  domain $\mathcal{D}$. We will further assume that the covariance between any
  two observations $y_i$ and $y_j$ located at $\mathbf{s}_i$ and $\mathbf{s}_j$
  is a function of only the distance $\|\mathbf{s}_i - \mathbf{s}_j\|$. Then the
  likelihood for $n$ observations from a single realization of $Y(\mathbf{s})$ is
  \begin{align}
    \ell_n(\boldsymbol{\theta}; \mathbf{y}_n) = &
      -\frac{n}{2}\log(2\pi)
      -\frac{1}{2}\log(|\boldsymbol{\Sigma}_n(\boldsymbol{\theta})|)
      \nonumber \\
      & - \frac{1}{2}\mathbf{y}_n'
                      \boldsymbol{\Sigma}_n(\boldsymbol{\theta})^{-1}
                      \mathbf{y}_n,
    \label{eqn:normal-loglikelihood-1}
  \end{align}  
  where
  $\boldsymbol{\Sigma}_{ij, n}(\boldsymbol{\theta}) =
  C(\boldsymbol{\theta}; \|\mathbf{s}_i - \mathbf{s}_j\|)$.
  
  While conceptually simple, these Gaussian process models
  present computational difficulties when the number of observations of the
  Gaussian process becomes large, as the likelihood function
  \eqref{eqn:normal-loglikelihood-1} requires the inversion of a $n \times n$
  matrix, which has computational cost $\mathcal{O}(n^3)$. To mitigate this
  cost, \citet{kaufman-2008b} proposed replacing
  \eqref{eqn:normal-loglikelihood-1} with the \emph{tapered} likelihood function
  \begin{align}
    \ell_{t,n}(\boldsymbol{\theta}; \mathbf{y}_n)  = &
      -\frac{n}{2}\log(2\pi)
      -\frac{1}{2}\log(|\boldsymbol{\Sigma}_n(\boldsymbol{\theta}) \circ \mathbf{T}_n|)
      \nonumber \\
      &
      -\frac{1}{2}\mathbf{y}_n'
                      \big(
                      (\boldsymbol{\Sigma}_n(\boldsymbol{\theta}) \circ \mathbf{T}_n)^{-1} 
                         \circ \mathbf{T}_n
                      \big)
                      \mathbf{y}_n,
    \label{eqn:tapered-loglikelihood}
  \end{align}
  where the $\circ$ notation denotes the element-wise product, and
  $\mathbf{T}_{ij} = \rho_{t}(\|\mathbf{s}_i - \mathbf{s}_j\|)$, a
  compactly-supported correlation function that takes a non-zero value when
  $\|\mathbf{s}_i - \mathbf{s}_j\|$ is less than some pre-specified ``taper
  range.'' The compact support of $\rho_{t}$ induces sparsity in $\mathbf{T}_n$,
  and hence all operations required to compute \eqref{eqn:tapered-loglikelihood}
  may be computed using specialized sparse-matrix algorithms, which are much
  faster and more memory-efficient than their dense-matrix analogues.
  
  Under suitable conditions, the tapered likelihood satisfies asymptotics of the
  form \eqref{eqn:asymp-theta-hat}, and Theorems \ref{thm:quasi-Bayes-normal}
  and \ref{thm:quasi-posterior-convergence} apply \citep{shaby-2012a}. For the
  simulations, we take $C(\boldsymbol{\theta}; \|\mathbf{s}_i - \mathbf{s}_j\|)
  = \sigma^2\exp\{-c/\sigma^2 \cdot \|\mathbf{s}_i - \mathbf{s}_j\|\}$, with
  $\mathbf{\theta} = (\sigma^2, c)' = (1, 0.2)'$. The observations are made on a
  $40 \times 40$ unit grid, so that each dataset $\mathbf{y}$ is a single
  1600-dimensional realization of a stochastic process. Half-Cauchy priors were
  used for both parameters.
  
  For this example, analytical expressions for both
  $\mathbf{P}(\boldsymbol{\theta})$ and $\mathbf{Q}(\boldsymbol{\theta})$ are
  available \citep{shaby-2012a}. As described in Section
  \ref{sec:estimating-Omega}, we use the plug-in estimator $\hat{\Omega}_k =
  \hat{\mathbf{Q}}_\text{III}(\hat{\boldsymbol{\theta}}_{\text{QB},k})^{-1}
  \hat{\mathbf{P}}_\text{II}(\hat{\boldsymbol{\theta}}_{\text{QB},k})^{1/2}
  \hat{\mathbf{Q}}_\text{III}(\hat{\boldsymbol{\theta}}_{\text{QB},k})^{1/2}$,
  as well as $\hat{\Omega}_k = \hat{\mathbf{Q}}_\text{I}^{-1}
  \hat{\mathbf{P}}_\text{II}(\hat{\boldsymbol{\theta}}_{\text{QB},k})^{1/2}
  \hat{\mathbf{Q}}_\text{I}^{-1}$, with $\hat{\mathbf{Q}}_\text{I}^{-1}$
  computed directly from the MCMC sample, for each $k=1, \ldots, 1000$ simulated
  datasets.

  \begin{figure}[ht]
    \begin{center}
      \begin{minipage}[t]{5cm}
        \includegraphics[width=\linewidth,
                         clip=true, trim=5 5 20 10]
                         {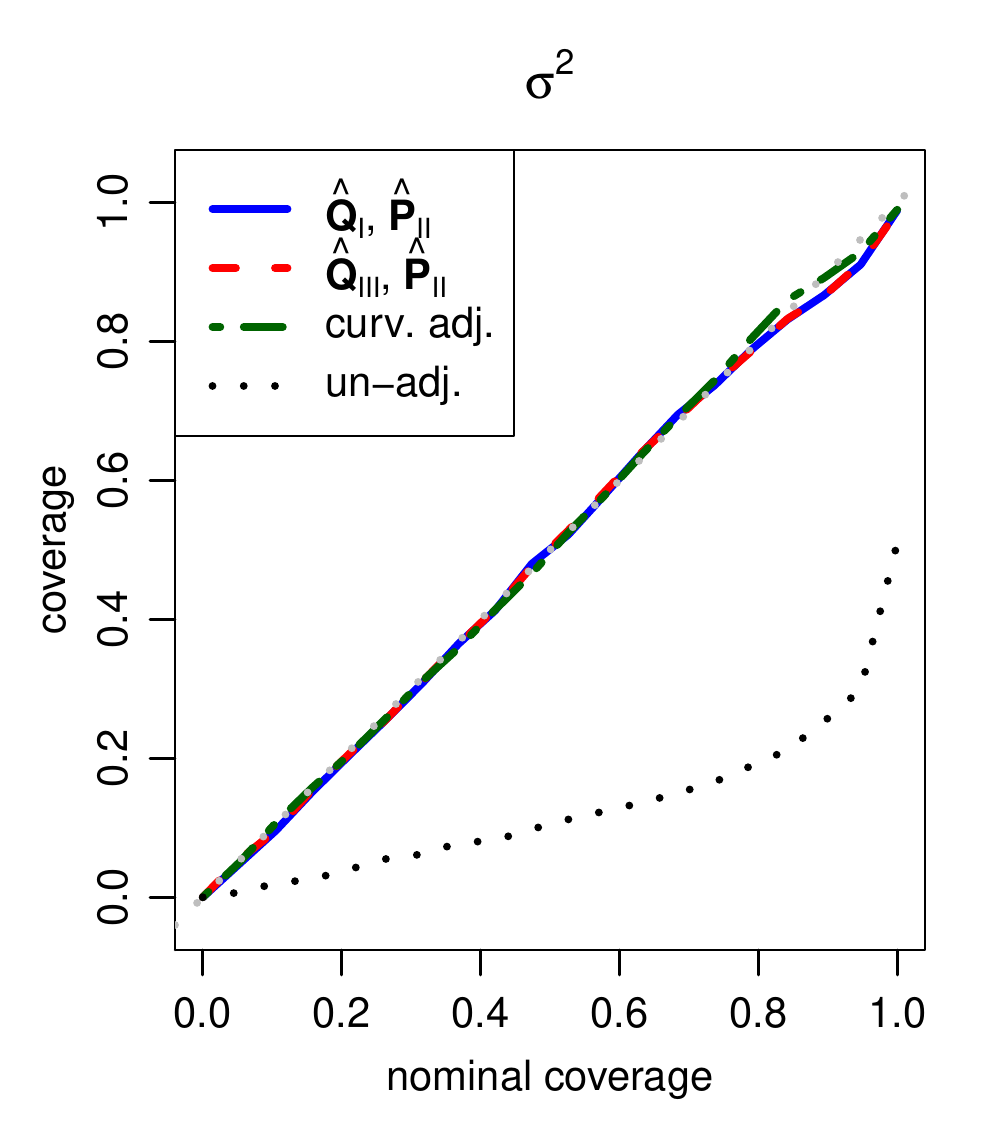}
      \end{minipage}
      \begin{minipage}[t]{5cm}
        \includegraphics[width=\linewidth,
                         clip=true, trim=5 5 20 10]
                        {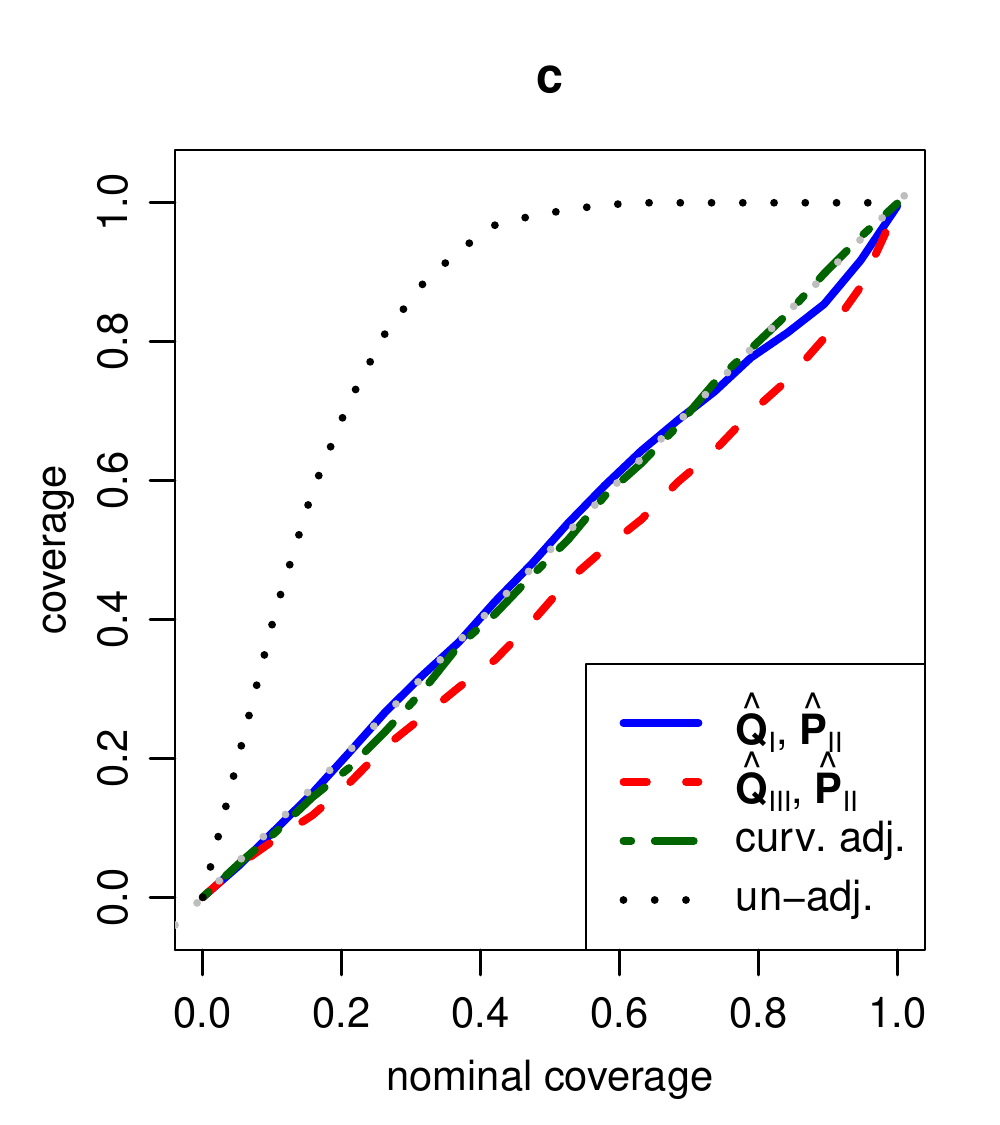}
       \end{minipage}
    \end{center}
    \caption{Empirical coverage rates for equi-tailed credible intervals based
             on MCMC samples using the tapered likelihood. Blue and red curves are
             OFS-adjusted samples using different estimates of $\bOmega$, green
             curves are from a curvature-adjusted sampler, and dotted curves are
             un-adjusted samples.}
    \label{fig:tapered-coverage}
  \end{figure}

  Figure \ref{fig:tapered-coverage} shows that the un-adjusted MCMC samples
  (dotted curves) yield horrible coverage properties for both $\sigma^2$ and
  $c$. It is somewhat interesting that while the ``naive'' intervals severely
  under-cover $\sigma^2$, they severely over-cover $c$. We therefore see that a
  naive implementation results in being overly optimistic about estimates of
  $\sigma^2$ while being overly pessimistic about estimates of $c$. The
  OFS-adjusted intervals display much more accurate coverage, achieving nearly
  nominal rates, although for $c$, the asymptotic expression for
  $\hat{\mathbf{Q}}$ seems to produce intervals that are systematically slightly
  too short. The curvature-adjusted sampler results in simliar coverage.
  
  To explore how the marginal OFS adjustment differs from the conditional
  adjustment in the Gibbs sampler setting, we simulate data from a spatial
  linear model, $Y(\mathbf{s}) \sim \text{GP}(\mathbf{X}\bbeta, C(\btheta);
  \mathbf{s})$. We set $\bbeta = (-0.5, 0, 0.5)'$ and use the same spatial
  design and covariance function as above. The design matrix $\mathbf{X}$ is a
  $1600 \times 3$ matrix of standard normal deviates, and the prior distribution
  for $\bbeta$ is a vague normal centered at zero. At each Gibbs iteration,
  $\btheta$ is updated using a Metropolis step using the tapered likelihood,
  and $\bbeta$ is updated by drawing directly from its conditionally conjugate
  full conditional distribution. The sampler is first run without adjustment,
  and $\hat{\btheta}_\text{QB}$ and $\hat{\bbeta}_\text{QB}$ are computed as
  the quasi-posterior means. The marginal OFS adjustment matrix is then
  computed using the analytical expression from \citet{shaby-2012a} by plugging
  in $\hat{\btheta}_\text{QB}$ and $\hat{\bbeta}_\text{QB}$. Next, the Gibbs
  sampler is run a second time using the estimated marginal OFS matrix to
  adjust the sample from the full conditional distribution of $\btheta$ at each
  iteration. Finally, the Gibbs sampler is run a third time, this time
  estimating the conditional OFS adjustment matrix at each iteration by
  maximizing the tapered likelihood function and plugging the resultant
  parameter estimates into the asymptotic formula for $\hat{\bOmega}$.
  
\begin{figure} [ht]
    \begin{center}
      \subfigure[]{\label{fig:cond_comparison-a}\includegraphics[width=.45\linewidth]
                                        {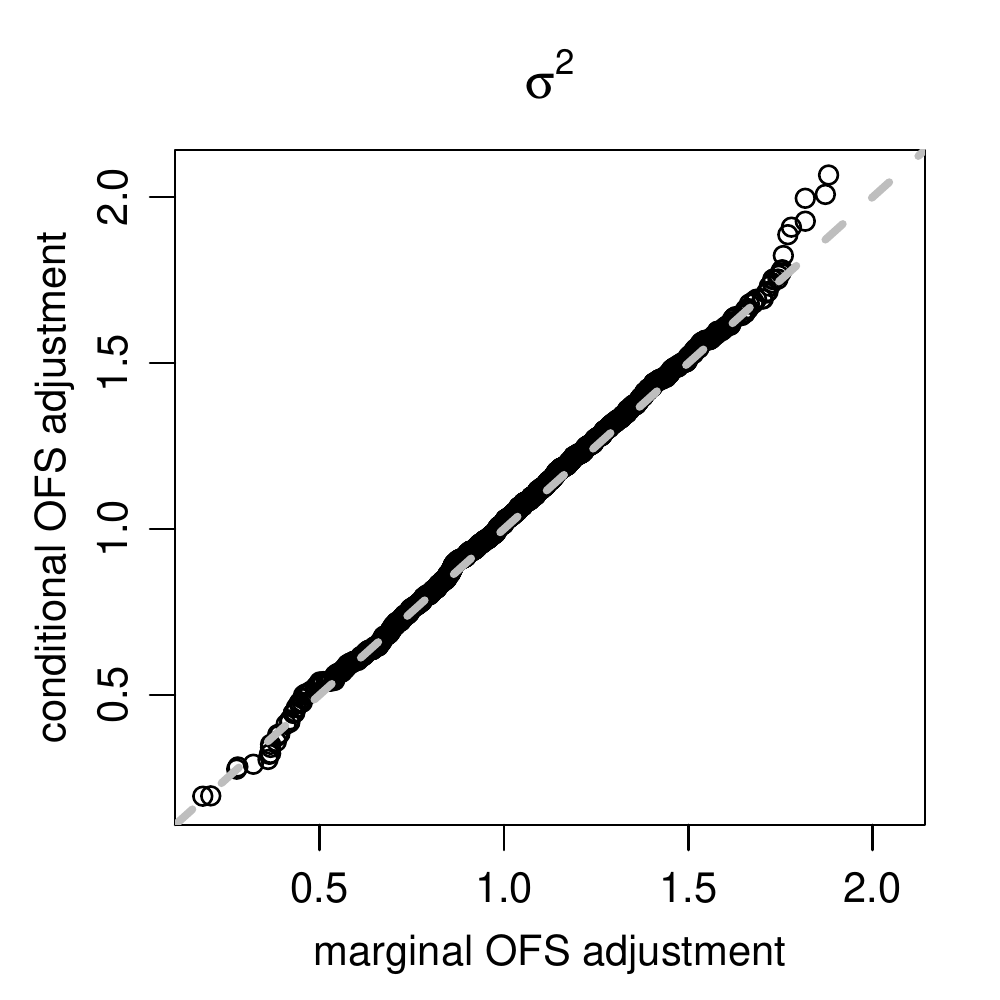}}
      \subfigure[]{\label{fig:cond_comparison-b}\includegraphics[width=.45\linewidth]
                                        {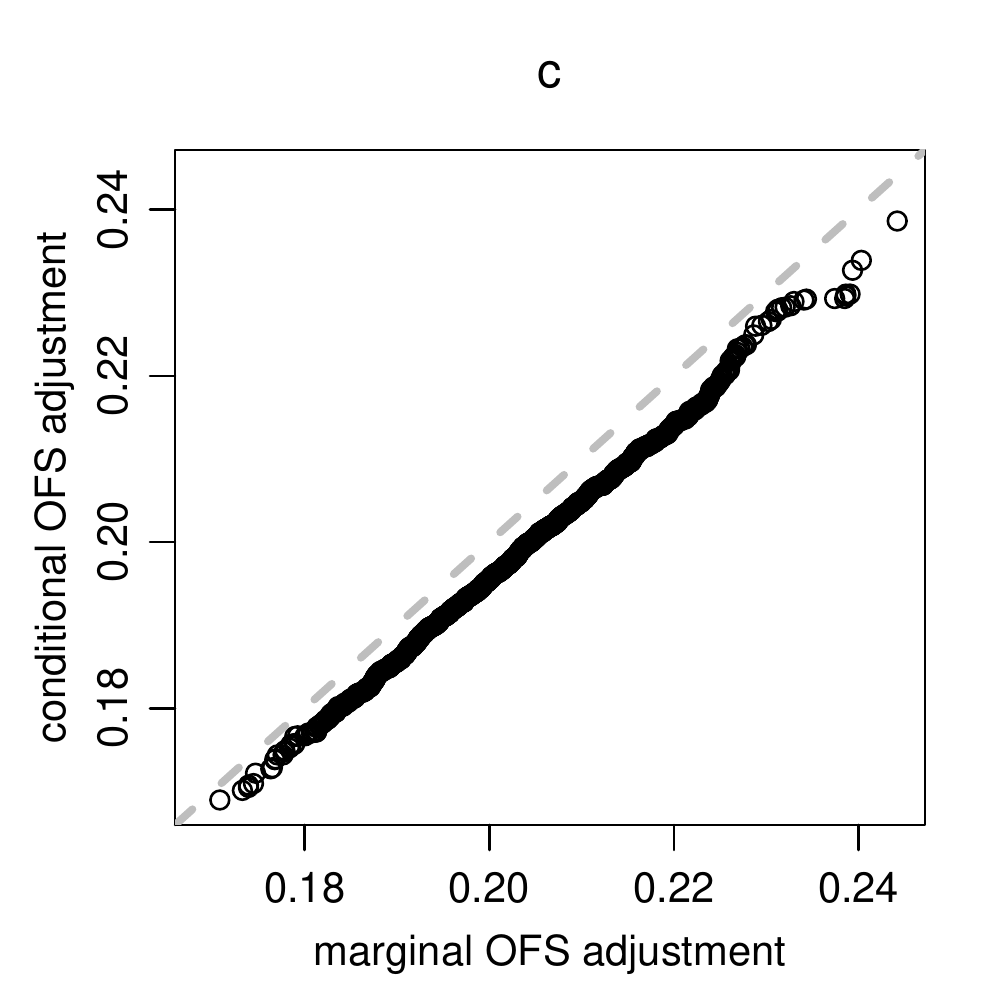}}
      \subfigure[]{\label{fig:cond_comparison-c}\includegraphics[width=.45\linewidth]
                                        {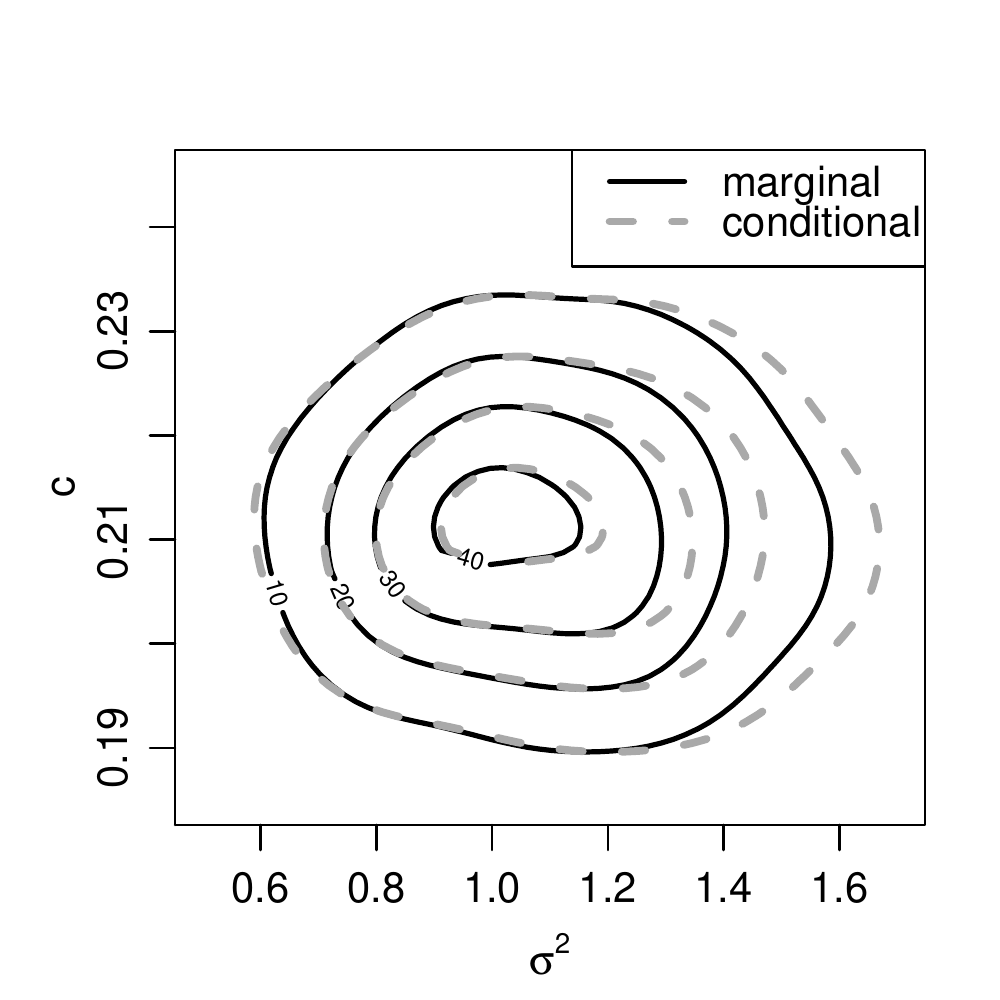}}
    \end{center}
    \caption{Comparison of marginal and conditional OFS-adjusted quasi-posterior
             distributions. Panels (a) and (b) show qq-plots of the marginal
             quasi-posteriors of the two covariance parameters. Panel (c) shows
             contour plots of a kernel density estimate of the joint
             quasi-posterior of the same parameters.} 
    \label{fig:compare_conditional}
  \end{figure}

  Because the conditional OFS-adjusted Gibbs sampler is so computationally
  expensive, we simulate just a few datasets and report the output from one of
  them. Figures \ref{fig:cond_comparison-a} and \ref{fig:cond_comparison-b}
  compare the marginal adjusted quasi-posterior distributions for the two
  covariance parameters, generated with the marginal and conditional OFS
  adjustments. The qq-plot for $\sigma^2$ shows almost perfect agreement except
  for a handful of MCMC samples on the upper tail. For the $c$ parameter, the
  qq-plot shows that the marginal OFS adjustment produces a quasi-posterior that
  is the same shape as that of the conditional adjustment, but is slightly
  shifted to the right. Figure \ref{fig:cond_comparison-c} shows contours of a
  kernel density estimate of the joint marginal ofs-adjusted quasi-posterior
  distribution of the two covariance parameters, with the marginal adjustment in
  black and the conditional adjustment in gray. The contours are very similar,
  with some small differences appearing on the right half of the $\sigma^2$-axis,
  indicating good agreement between the two bivariate distributions. The output
  from all the simulated datasets looked qualitatively similar, with no
  noticeable systematic differences between the two adjustments. The choice of
  adjustment had no discernible effect on the quasi-posterior distribution of
  $\bbeta$.

\subsection{Composite likelihood for max-stable processes}
\label{sec:composite}

  Statistical models for extreme values that include spatial dependence are
  useful for studying, for example, extreme weather events like heat waves and
  powerful storms \citep[][e.g.]{cooley-2007a, sang-2010a}. Extreme value theory
  tells us that the distribution of block-wise maximum values (such as annual
  high temperatures) of independent draws from any distribution converges to a
  generalized extreme value (GEV) distribution \citep[see][]{coles-2001a}, if it
  converges at all. The asymptotics therefore suggest that any model of
  block-wise maxima at several spatial locations ought to have GEV marginal
  distributions with distribution function
  \begin{displaymath}
    F(y; \mu, \sigma, \xi) = \exp\Big\{ -\Big[1+\xi\Big( \frac{x-\mu}{\sigma}
                                                   \Big)
                                         \Big]^{-1/\xi}  \Big\},
  \end{displaymath}
  where $\mu$ is a location, $\sigma$ a scale, and $\xi$ a shape parameter that
  determines the thickness of the right tail. The GEV may be characterized by
  the max-stability property \citep{coles-2001a}. More generally, block-wise
  maxima of random vectors must also converge to max-stable
  distributions.
  
  \citet{sang-2010a} achieve spatial dependence with GEV marginals through a
  Gaussian copula construction. However, Gaussian copula models have been
  strongly criticized \citep{Kluppelberg-2006a} because they do not result in
  max-stable finite-dimensional distributions, nor do they permit dependence in
  the most extreme values, referred to as \emph{tail dependence}, and it is
  clear that physical phenomena of interest do exhibit strong spatial dependence
  even among the most extreme events.
  
  An alternate approach for encoding spatial dependence of extreme values is
  through max-stable process models \citep{dehaan-1984a}, which are stochastic
  processes over some index set where all finite-dimensional distributions are
  max-stable. Explicit specifications of spatial max-stable processes based on
  the \citet{dehaan-1984a} spectral representation have been proposed by
  \citet{smith-1990a}, \citet{schlather-2002a}, and \citet{kabluchko-2009a}.
  These formulations have the advantage that they do represent tail dependence.

  Unfortunately, for all of the available spatial max-stable process models,
  joint density functions of observations at three or more spatial locations are
  not known (a slight exception is the Gaussian extreme value process (GEVP)
  \citep{smith-1990a}, for which \citet{genton-2011a} derives trivariate
  densities). Since bivariate densities are known, \citet{padoan-2010a} proposes
  parameter estimation and inference via the \emph{pairwise likelihood}, where
  all bivariate log likelihoods are summed as though they were independent:
  \begin{displaymath}
    \ell_p(\boldsymbol{\theta}; \mathbf{y}) = 
    \sum_{i \neq j}\ell(y_i, y_j; \boldsymbol{\theta}).
  \end{displaymath}
  The pairwise likelihood is a special case of a composite likelihood
  \citep{lindsay-1988a}. \citet{padoan-2010a} show that asymptotic normality of
  the form \eqref{eqn:asymp-theta-hat} applies, so we may again apply the OFS
  adjustment.
  
\begin{figure}[ht]
    \begin{center}
      \includegraphics[width=8cm,
                       clip=true, trim=110 120 110 110]
                       {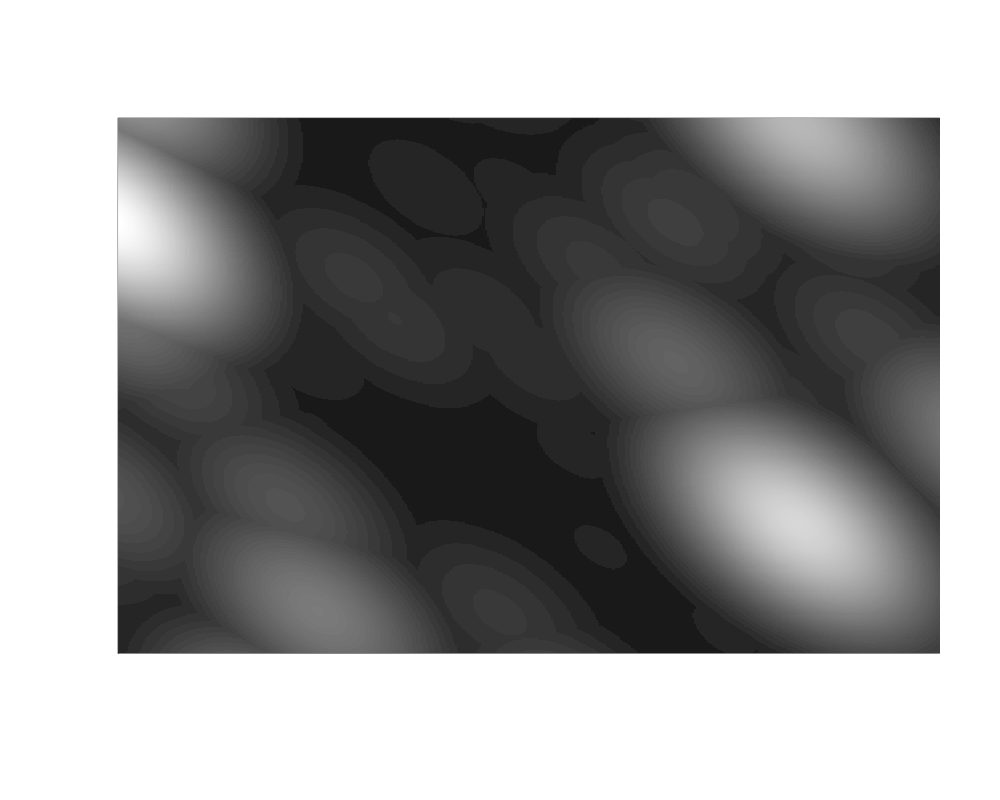}
    \end{center}
    \caption{A realization of the Gaussian extreme value max-stable process of
             \citet{smith-1990a}.}
    \label{fig:smithproc}
  \end{figure}

  Our simulation experiment consists of 1000 draws from a GEVP
  with unit Fr\'echet margins on a $10 \times 10$ square grid, with
  100 replicates per draw. An example of a single replicate is shown in Figure
  \ref{fig:smithproc}. This setup would correspond, for example, to 100 years of
  annual maximum temperature data from 100 weather stations.

  The unknown parameter $\boldsymbol{\theta}$ in a 2-dimensional
  GEVP is a $2 \times 2$ covariance matrix. For this
  simulation, $\boldsymbol{\theta}_0 = (\Sigma_{11}, \Sigma_{12}, \Sigma_{22})'
  = (0.75, -0.5, 1.25)'$. The prior distribution for $\boldsymbol{\Sigma}$ is a
  vague inverse-Wishart. For each draw, a long MCMC chain is run and
  $\hat{\boldsymbol{\theta}}$ is computed as the posterior mean. In addition,
  for each draw, all four $\mathbf{Q-P}$ combinations of
  $\hat{\mathbf{Q}}_\text{I}, \hat{\mathbf{Q}}_\text{II},
  \hat{\mathbf{P}}_\text{I}$, and $\hat{\mathbf{P}}_\text{boot}$, as defined
  above, are computed to produce four estimates of $\bOmega$. Finally, the
  curvature-adjusted MCMC sampler from \citet{ribatet-2012a} is run on each
  simulated dataset, with $\bOmega$ estimated from $\hat{\mathbf{Q}}_\text{II}$
  and $\hat{\mathbf{P}}_\text{I}$, evaluated at the maximum pairwise likelihood
  estimate of $\bSigma$.
  

  \begin{figure} [ht]
    \begin{center}
      \subfigure[]{\label{fig:smithproc-coverage-a}
                   \includegraphics[width=.45\linewidth,
                                    clip=true, trim=5 5 20 10]
                                   {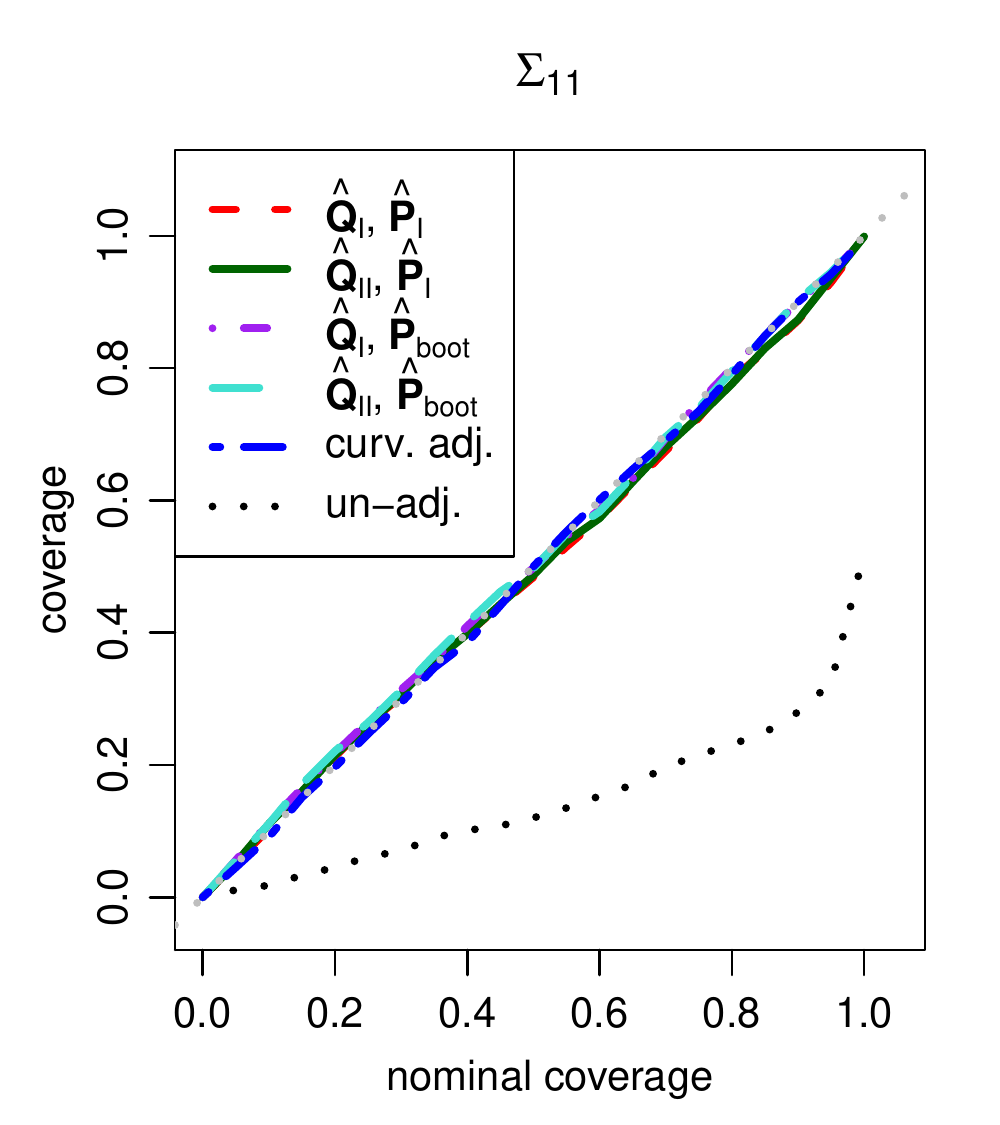}}
      \subfigure[]{\label{fig:smithproc-coverage-b}
                   \includegraphics[width=.45\linewidth,
                                    clip=true, trim=5 5 20 10]
                                  {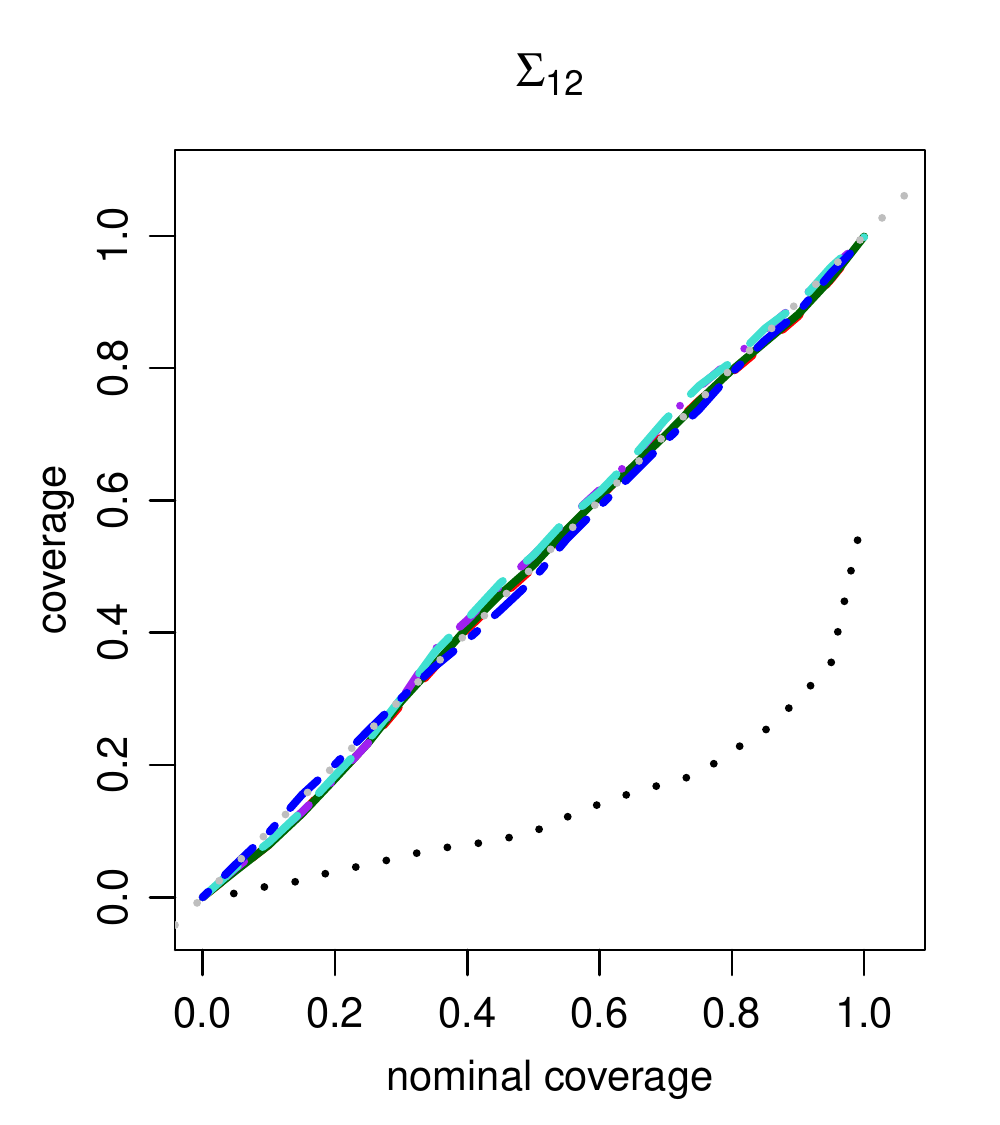}}
      \subfigure[]{\label{fig:smithproc-coverage-c}
                   \includegraphics[width=.45\linewidth,
                                    clip=true, trim=5 5 20 10]
                                   {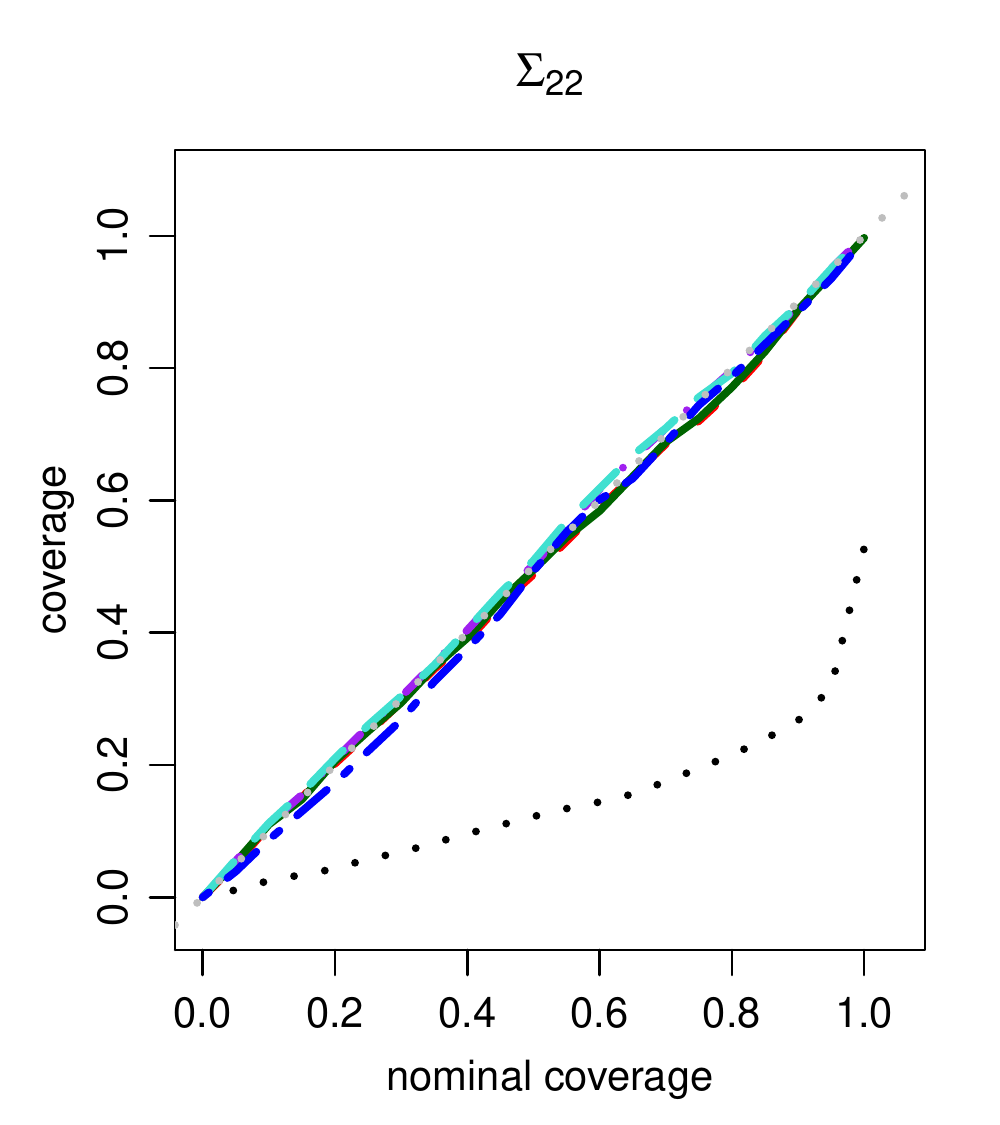}}
    \end{center}
    \caption{Empirical coverage rates for equi-tailed credible intervals based
             on MCMC samples using the pairwise likelihood. Colored curves are
             OFS-adjusted samples using different estimates of $\bOmega$, and
             from a curvature-adjusted sampler. Dotted curves are un-adjusted
             samples.}
    \label{fig:smithproc-coverage}
  \end{figure}

  Figure \ref{fig:smithproc-coverage} shows that, for this simulation, the OFS
  adjustment produces credible intervals that cover at almost exactly their
  nominal rates. Furthermore, OFS-adjusted credible intervals based on the four
  values of $\hat{\boldsymbol{\Omega}}$ turned out nearly identical. This is
  about as good a result as one could hope for. The curvature-adjusted sampler
  also achieves nominal coverage. The un-adjusted intervals systematically
  under-cover for each of the three parameters. This is expected (as noted by
  \citet{ribatet-2012a}), as the pairwise likelihood over-uses the data by
  including each location in roughly $n/2$ terms of the objective function rather
  than just one, as would be the case with a likelihood function. This results in
  a pairwise likelihood surface that is far too sharply-peaked relative to a
  likelihood surface. The OFS adjustment seems to successfully compensate for
  this effect.

\section{Data analysis}
\label{sec:data}

\subsection{Bird Counts}
\label{sec:bird-counts}

  Next, we apply tapered quasi-Bayesian analysis to a hierarchical
  model that includes a Gaussian process. Instead of a purely spatial random
  field as in Section \ref{sec:tapering}, we assume a spatio-temporal random
  field, which highlights some of the advantages of tapering over other methods
  designed for large spatial datasets.

  The dataset comes from a ``citizen science'' initiative called eBird
  (\url{www.ebird.org}). The idea of citizen science is that many
  non-professional observers can be leveraged to collect an enormous amount of
  data. eBird participants across North Americal record the birds they see,
  along with the time and location of the observation, into a web-based
  database. Here, we look at 6114 observations of the Northern Cardinal in a
  section of the eastern United States over a period from 2004 to 2007 (Figure
  \ref{fig:norcar-locations}).

  Inspection of the data suggests an overdispersed Poisson model.  Let
  $Y_i, \ldots, Y_n$ be observed counts and
  $\mathbf{X=[x}_1,\ldots,\mathbf{x}_n]$ be a matrix of covariates
  associated with each observation.  Also, let $\mathbf{S=[s}_1,
  \ldots, \mathbf{s}_n]$ and $\mathbf{T=[t}_1, \ldots, \mathbf{t}_n]$
  be the spatial and temporal locations, respectively, associated with
  $Y_i, \ldots, Y_n$, with space indexed by latitude and longitude.

  For this example, we deliberately chose a small number of predictors, several
  of which vary spatially. Preliminary analyses led to a set of 10 covariates
  that includes time of day, day of year, human population density, percentage
  of developed open space (single-family houses, parks, golf courses, etc.),
  tree canopy density, and variables that measure observer effort. Simple
  transformations (logs, powers, etc.) were applied to some of the covariates,
  as suggested by ornithologists and exploratory analyses.

  We specify the model as
  \begin{eqnarray}
    y_i | \lambda_i & \sim & \mbox{Pois}(\lambda_i) \nonumber   \\
    \log{\lambda_i}(\mathbf{x}_i, \mathbf{s}_i, \mathbf{t}_i)    & = &
    \mathbf{x}'_i\boldsymbol{\beta} + Z_i(\mathbf{s}_i, \mathbf{t}_i)
    + \varepsilon_i 
     \nonumber \\
     \mathbf{z}(\mathbf{S, T})|\boldsymbol{\theta}^*          & \sim &
    N(0, \boldsymbol{\Sigma}^*(\boldsymbol{\theta}^*;
                                     \mathbf{S, T}))  \nonumber \\
    \varepsilon_i | \tau                                 & \sim &
    \mbox{iid } N(0, \sigma^2_\varepsilon).
    \label{mod:poisson-1}
  \end{eqnarray}
  We assume that the random effect $\mathbf{z(S, T)}$ has a Gaussian
  random field structure.  Thus, this model is an example of
  ``model-based geostatistics'' of \citet{diggle-1998a}, a class of spatial
  generalized linear models that has seen wide application in the environmental
  literature.

  Even though Northern Cardinals are not migratory birds, a
  spatio-temporal structure for the random effect has a great
  intuitive appeal.  One can easily imagine clusters of birds habiting
  different locales, moving from place to place based on things like
  food availability or disturbances. The correlation range of the
  spatio-temporal random effect informs the ornithologists about the scales of
  movements of Northern Cardinals in space and time, as well as providing clues
  about what un-measured covariates are needed to explain the pattern of
  Northern Cardinal observations.

  The parameter $\boldsymbol{\varepsilon}$ can be interpreted as either an
  overdispersion parameter, or as the traditional ``nugget'' effect,
  representing small-scale variation or measurement error. It will be convenient
  to marginalize over the random effects $\mathbf{z}$ and
  $\boldsymbol{\varepsilon}$ and consider the distribution of the log-means
  directly. Furthermore, we will write the matrix
  $\boldsymbol{\Sigma}^*(\boldsymbol{\theta}^*;\mathbf{S, T}) +
  \sigma^2_\varepsilon\mathbf{I}$ simply as
  $\boldsymbol{\Sigma}(\boldsymbol{\theta};\mathbf{S, T})$ and condense
  $\boldsymbol{\theta}^*$ and $\sigma^2_\varepsilon$ into the single parameter
  vector $\boldsymbol{\theta}$. The resulting model, equivalent to
  \eqref{mod:poisson-1}, is written as
  \begin{eqnarray}
    y_i | \lambda_i   & \sim & \mbox{Pois}(\lambda_i)   \nonumber  \\
    \log{\lambda_i}(\mathbf{x}_i, \mathbf{s}_i, \mathbf{t}_i)   
     & \equiv & b_i     \nonumber \\
    \mathbf{b}(\mathbf{X}, \mathbf{S, T}) | 
      \boldsymbol{\theta}, \boldsymbol{\beta}  & \sim &
    N(\mathbf{X}\boldsymbol{\beta},
      \boldsymbol{\Sigma}(\boldsymbol{\theta}; \mathbf{S, T})).
    \label{mod:poisson-2}
  \end{eqnarray}

  Another level in the hierarchy imposes a ridge penalty on the
  regression coefficients $\boldsymbol{\beta}$, specified as
  
  \begin{equation}
    \boldsymbol{\beta} \sim N(0, \sigma^2_\beta \mathbf{I}).
    \label{mod:ridge-beta}
  \end{equation}
  Finally, we need priors for the parameters $\boldsymbol{\theta}$ and
  $\sigma^2_\beta$
  \begin{eqnarray}
    \theta_i & \sim & \pi_{\theta_i} \nonumber \\
    \sigma^2_\beta  & \sim & \pi_{\sigma^2_\beta}      \nonumber 
    \label{mod:priors-1}
  \end{eqnarray}
  independently.  

  For $\boldsymbol{\Sigma}(\boldsymbol{\theta}; \mathbf{S, T})$, we
  chose a spatio-temporal covariance model from
  \citet{gneiting-2002a}.  The covariance functions described therein
  are nonseparable in that (except in special cases) they cannot be
  written as the product of a purely spatial and purely temporal
  covariance function.  Specifically, we let

  \begin{equation}
    C(\boldsymbol{\theta}^*; \mathbf{h, u}) =
    \frac{\sigma^2}{(a\mathbf{u}^{2\alpha} + 1)^2} \cdot
    \exp{\bigg\{\frac{-(c/\sigma^2)\mathbf{h}^{2\gamma}}
                     {(a\mathbf{u}^{2\alpha} + 1)^{\omega\gamma}}  \bigg\}},
    \label{eqn:gneiting-1}
  \end{equation}
  where $\mathbf{h}$ and $\mathbf{u}$ are distances between
  observation points in space and time, respectively.  The parameters
  $\alpha \in (0, 1]$ and $\gamma \in (0, 1]$ control the smoothness
  of the process.  We fixed these parameters at convenient values of 1 and .5,
  respectively, because they were not well-identified by the data.  

  The parameter $\omega \in [0, 1]$ has the nice interpretation of
  specifying the degree of nonseparability between purely spatial and
  purely temporal components; when $\omega = 0$,
  $C(\boldsymbol{\theta}; \mathbf{h, u})$ is the product of a purely
  temporal and a purely spatial (exponential) covariance function.

  Priors for the parameters $\sigma^2$, $a$, $c$, $\sigma^2_\varepsilon$, and
  $\sigma^2_\beta$ are specified as vague Cauchy distributions,
  truncated to have only positive support.  The interaction parameter
  $\omega$ is given a uniform prior on $[0, 1]$.

  A valid spatio-temporal taper matrix may be constructed as the
  element-wise product of a spatial and a temporal taper matrix

  \begin{displaymath}
    \mathbf{T} = \mathbf{T}_s \circ \mathbf{T}_t.
  \end{displaymath}
  Constructed this way, $\mathbf{T}$ inherits the sparse entries of
  both $\mathbf{T}_s$ and $\mathbf{T}_t$, and may therefore itself be
  extremely sparse.

  Several other methods exist to mitigate the computational burden imposed by
  large spatial datasets with non-Gaussian responses. \citet{wikle-2002a} and
  \citet{royle-2005a} embed a continuous spatial process into a
  latent grid and work in the spectral domain using fast Fourier methods.
  However, applying Fourier methods here is problematic, as it is not obvious
  how to do so for a process that has spatio-temporal structures. Low rank
  methods like predictive processes \citep{banerjee-2008a, finley-2009a} and
  fixed rank Kriging \citep{cressie-2008a} are also popular for spatial data.
  Applying these methods to spatio-temporal models is possible, but awkward. For
  predictive processes, one must decide how to specify knot locations in
  space $\times$ time. For fixed rank Kriging, one must specify knot locations
  as well as space-time kernel functions. Fixed rank Kriging as been adapted to
  the spatio-temporal setting \citep{cressie-2010a} through a linear filtering
  framework, but only for Gaussian responses. Finally, Gauss-Markov
  approximations to continuous spatial processes are fast to compute, especially
  when using Laplace approximations in place of MCMC \citep{lindgren-2011a,
  rue-2009a}. However, again, these methods do not apply to data with
  spatio-temporal random effects.

  In contrast, application of the tapering approach in the spatio-temporal
  context is immediate and even potentially enjoys increased computational
  efficiency relative to the purely spatial context because of the additional
  sparsity induced by element-wise multiplication with the temporal taper matrix.  

  For the eBird data, a taper range of 20 miles and 60 days gives a tapered
  covariance matrix with about .5\% nonzero elements. MCMC was carried out using
  a block Gibbs sampler. Each evaluation of the expensive normal log likelihood
  was replaced by its tapered analogue. Within each Gibbs iteration, each of
  $\mathbf{b}$, $\boldsymbol{\theta}$, and $\sigma^2_\beta$ are updated with a
  random walk Metropolis step. The full conditional
  distribution for $\boldsymbol{\beta}$ is conditionally conjugate, enabling a
  simple update as a draw from the appropriate normal distribution.

  As described in Section \ref{sec:ofs-gibbs}, the tapered Gibbs sampler was run
  twice. Samples from the first run were used to produce point estimates of
  $\btheta$ and the marginal OFS adjustment matrix $\bOmega_{\btheta | \bbeta,
  \mathbf{b}, \sigma^2_\beta}$. The estimate $\hat{\bOmega}_{\btheta | \bbeta,
  \mathbf{b}, \sigma^2_\beta}$ was computed from the asymptotic expressions for
  $\mathbf{Q}$ and $\mathbf{Q}$ evaluated at $\hat{\btheta}_\text{QB}$, the
  quasi-posterior mean. Because the quasi-posterior distribution of interaction
  parameter $\omega$ was nearly uniform on [0,1], it was excluded from the
  adjustment. The conditional OFS was not attempted, as doing so would have
  required several months of computation time.

\begin{figure} [ht]
    \begin{center}
      \includegraphics[width=\linewidth, angle=0,
                       clip=true, trim=0 0 0 5]
                      {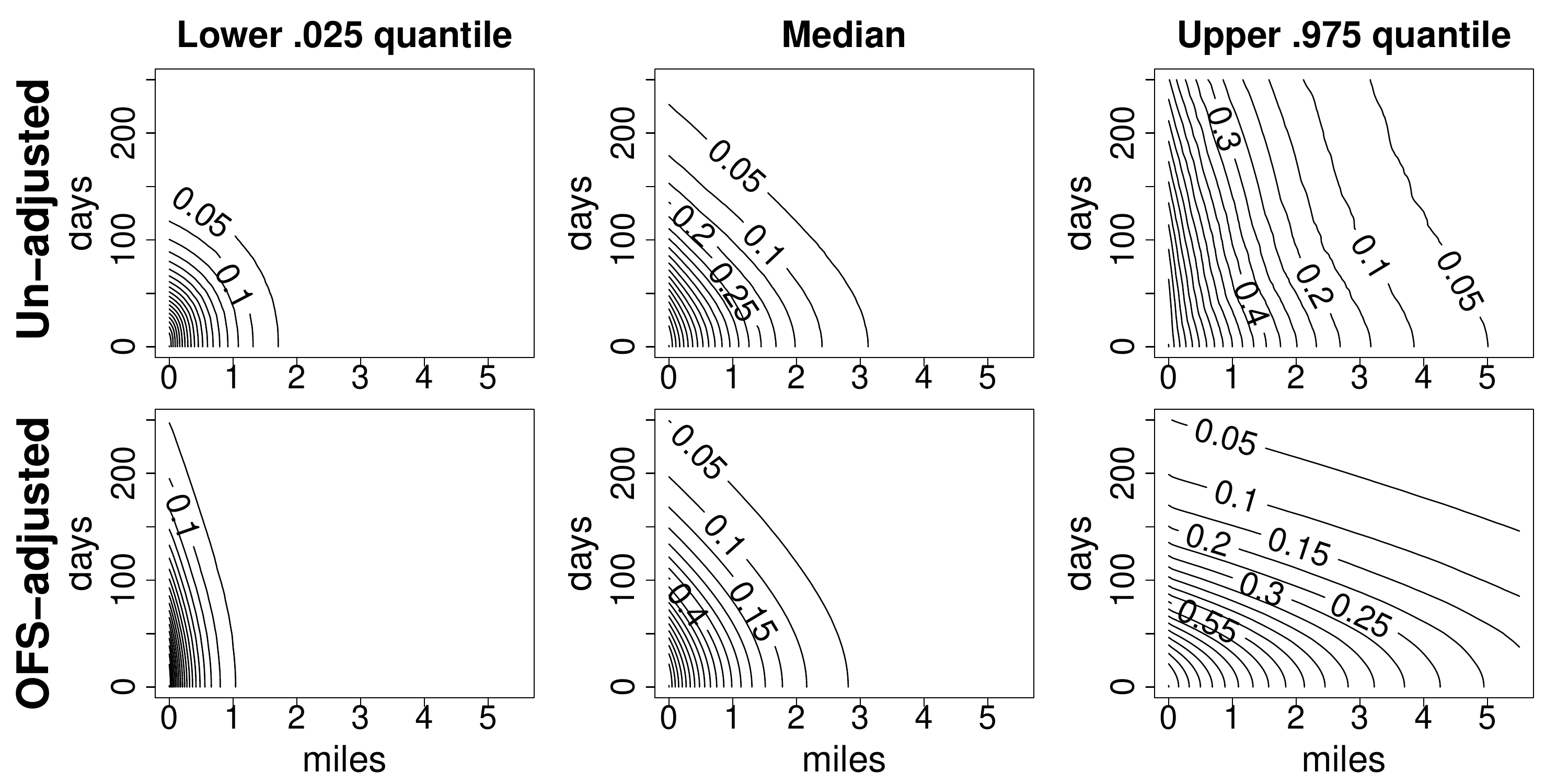}
    \end{center}
    \caption[]{Posterior quantiles of spatio-temporal correlation surface. The
               top row was computed from the un-adjusted sampler, and the bottom
               row was computed from the OFS-adjusted sampler. The median
               surfaces are similar, but the high and low quantile contours show
               that the uncertainty is noticeably altered by the adjustment.}
    \label{fig:correlation-contours}
  \end{figure}

  After discarding 5000 burn-in iterations, 5000 MCMC samples were used
  for estimation and prediction. Pointwise quantiles of the posterior
  correlation surface are shown in Figure \ref{fig:correlation-contours}, for
  both the un-adjusted and adjusted samples. The point at which the correlation
  drops to .05, often called the ``effective range'' of a process, is the most
  extreme contour displayed in each of the plots in figure
  \ref{fig:correlation-contours}. While the two sets of contours do not differ
  much in the median, they are quite different in the upper and lower quantiles.
  For this analysis, the correlation structure is a key component with a useful
  interpretation, so its posterior uncertainty is of interest. Comparing the
  OFS-adjusted and un-adjusted correlation surfaces, it is interesting to note
  that OFS adjustment gives decreased temporal uncertainty but increased spatial
  uncertainty.
  
  The fairly long median effective range of around 225 days at spatial lag 0
  seemed reasonable to a panel of ornithologists, as Northern Cardinals, while
  they do move around to some degree, are not migratory birds. The effective
  range of 3 miles at time lag 0 seemed reasonable as well. Northern Cardinals
  build new nests each year and are socially monogamous within a breeding
  season, but divorces sometimes occur between years. They generally stay close
  to the nest to forage and bathe. Males are highly territorial and will
  occasionally challenge neighboring males' breeding territories. These
  behaviors are consistent with a the posterior median temporal dependence range
  of a significant fraction of a single year, and a posterior median spatial
  range that is larger than but in the ballpark of an individual's territorial range.

\begin{figure}[ht]
    \begin{center}
      \includegraphics[width=6cm, angle=0,
                       clip=true, trim=0 0 0 5]
                      {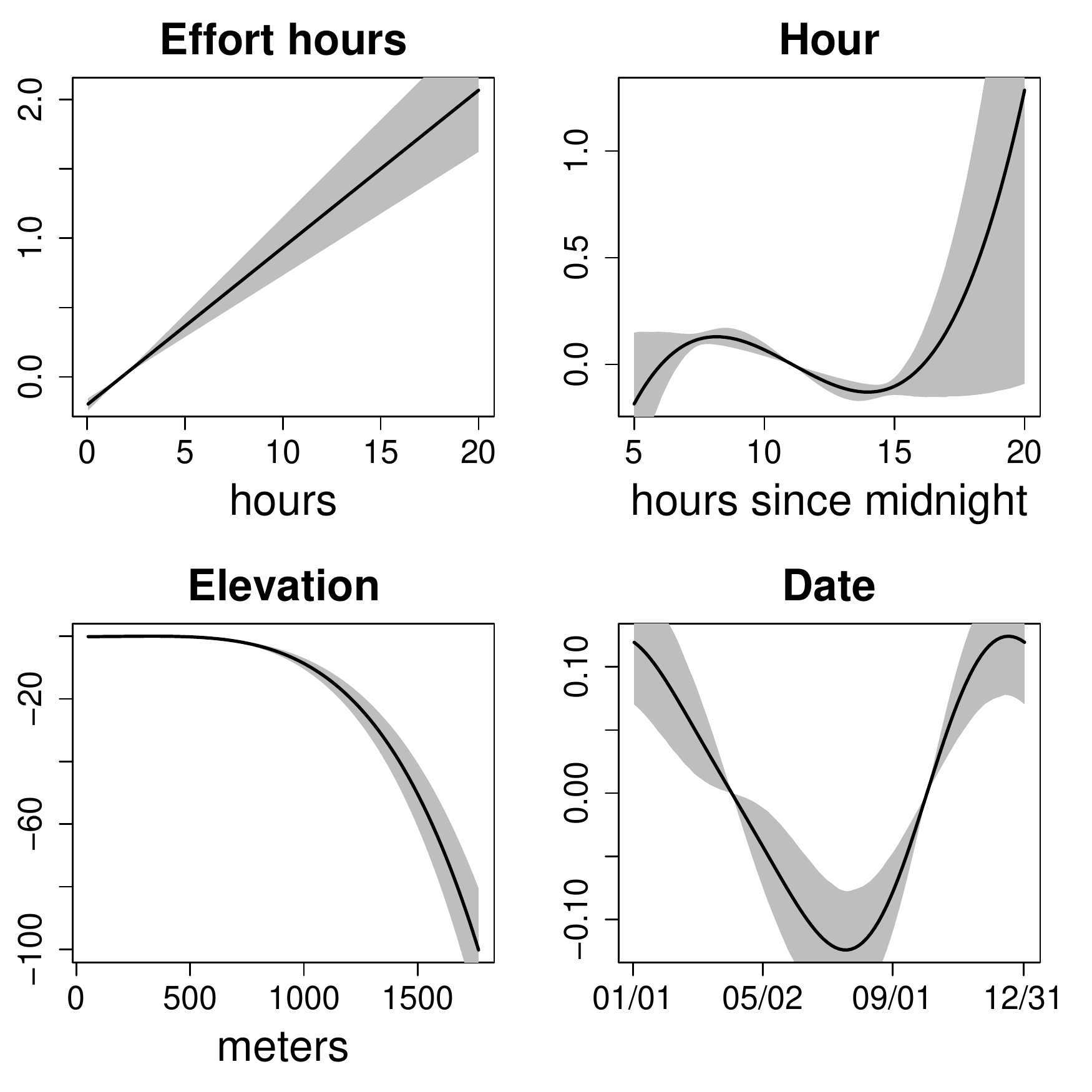}
    \end{center}
    \caption[]{A subset of the estimated fixed effects. Solid lines are
              posterior medians, and shaded areas are posterior pointwise 90\%
              credible intervals.}
    \label{fig:norcar-fixed-effects}
  \end{figure}

  Posterior estimates for some of the more interesting fixed effects,
  along with 95\% pointwise credible intervals, are plotted in Figure
  \ref{fig:norcar-fixed-effects}. The top right panel shows a clearly increasing
  trend as a function of the number of hours spent observing. The top right
  panel shows that the effect as a function of time of day increases until about
  8 a.m. and then decreases until about 3 p.m., when it again begins to
  increase. The increase after 3 p.m. is accompanied by very wide credible
  intervals. In the bottom left panel, we can see an overwhelming negative
  effect at high elevations. Finally, the bottom right panel shows a seasonal
  cycle that peaks in early winter and attains its minimum in later summer. 

  Recall that these fixed effects are on the log scale.  Here again, a
  panel of ornithologists was pleased with the results.  Obviously,
  the number of observed counts should increase with the amount of
  time an observer spends watching. The peak in the time of day effect
  at around 8 a.m. reflects the time of the highest activity level of
  the birds.  The wide confidence bands starting at around 4 p.m.
  probably results from a lack of data in the afternoon.  Northern
  Cardinals cannot live in habitats found at higher elevations, a fact
  reflected in the huge negative effect estimated after about 700
  meters.  Finally, cardinals tend to be easier to detect during the
  winter months because they are more vocal, and they visit feeders
  more frequently.  In the summer months, they tend to stay more
  hidden because it is their breeding season, and they do not visit
  feeders as often because food is more plentiful.  These seasonal
  variations in detectability are reflected in the pattern shown in
  the estimated date effect.

\begin{figure}[ht]
    \begin{center}
      \includegraphics[width=8cm, angle=0,
                       clip=true, trim=0 45 0 40]
                      {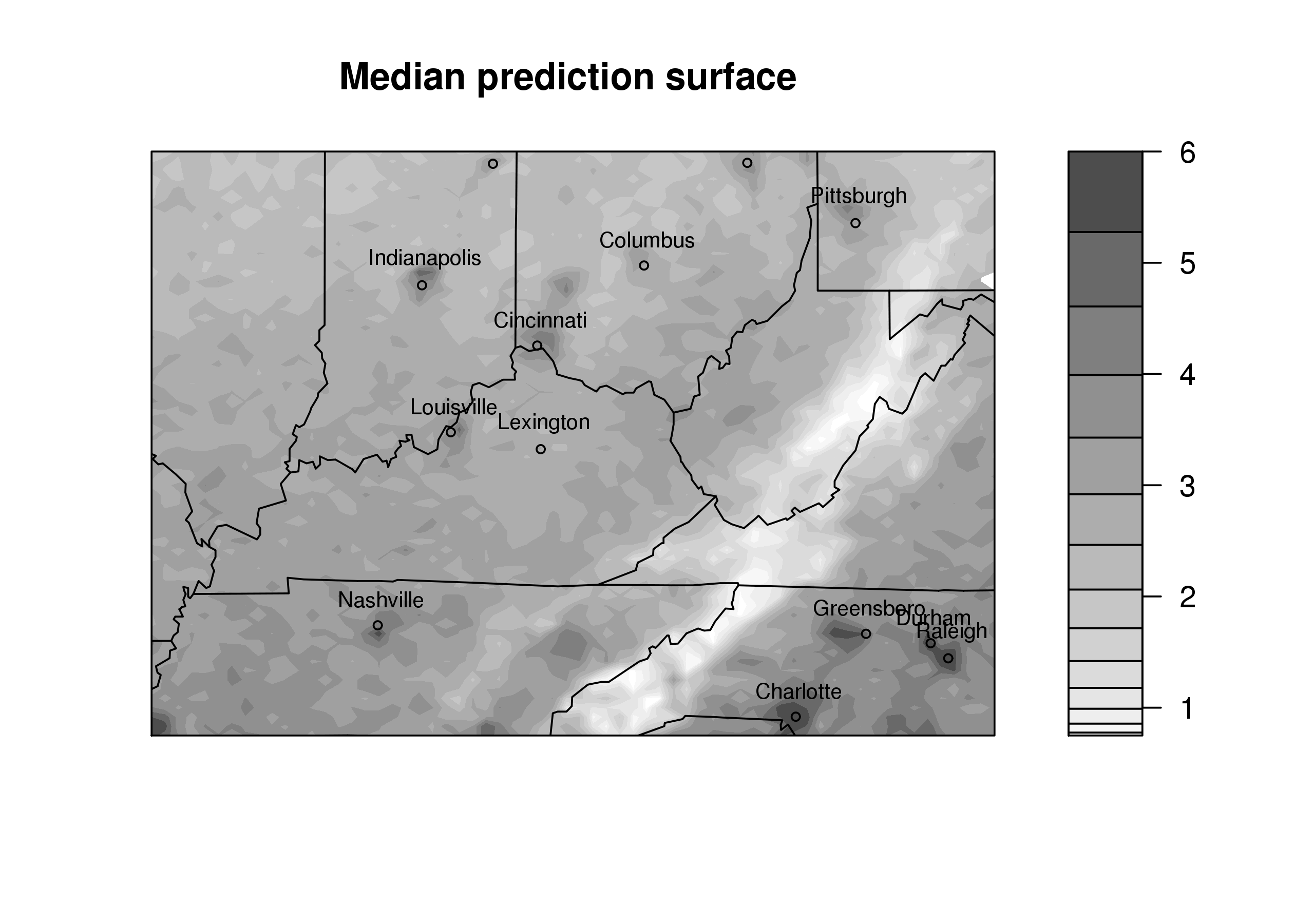}
    \end{center}
    \caption[]{Median predicted surface for 8 a.m. on April 11.}
    \label{fig:norcar-predicted-surface}
  \end{figure}

  The median posterior predicted surface (Figure
  \ref{fig:norcar-predicted-surface}) of the mean counts was
  generated by drawing from the posterior predictive distribution at a
  large set of sample points in the spatial domain, for fixed values
  of ``effort'' covariates, and at a fixed time.

  Maps like Figure \ref{fig:norcar-predicted-surface}, of course, vary
  in time as well as space.  The most prominent feature
  of the predicted surface is the very low values along the
  Appalachians.  This is a result of the huge elevation effect, and
  corroborates expert knowledge.  Another noticeable feature of the
  prediction surface is the elevated counts around population centers.
  It is well-known among ornithologists that Northern Cardinals are
  most common in the suburbs.  This happens for two reasons.  The
  first is that they are attracted to the many bird feeders found in
  the suburbs.  The second is that suburban habitat, with landscaped
  gardens and mixes of open areas, shrubs, and trees, is ideal habitat
  for cardinals.

\section{Discussion}
\label{sec:discussion}

  The open-faced sandwich adjustment provides a way to incorporate estimating
  functions that are not likelihoods into Bayesian-like models. While the
  resulting inference does not enjoy the elegant formal probabilistic
  interpretation of pure Bayesian analysis, it does inherit some of its most
  desirable attributes: borrowing strength across parameters, the ability to
  work with complicated hierarchical structures, and propagating uncertainty
  throughout model components, to name a few. When the likelihood function is
  unknown or has undesirable properties, the OFS adjustment allows one to retain
  these beneficial features of Bayesian analysis while avoiding the need to
  compute the likelihood function by substituting a suitable objective function
  in its place.
  
  These benefits come with a few costs. First, the resultant MCMC samples may
  not be interpreted as though they came from a true Bayesian model. Second, the
  coverage characteristics of the output are only as good as the
  applicability of the asymptotic approximation and the practitioner's ability
  to estimate the sandwich matrix, which can be a difficult task in some
  situations \citep{kauermann-2001a}. Third, estimating the adjustment matrix
  using sample moments or a bootstrap and relying on it to produce posterior
  samples has a decidedly ``un-Bayesian'' feel to it. Finally, using the
  adjustment in the Gibbs sampler context does require approximately twice the
  computational effort as the un-adjusted Gibbs sampler, which in some cases can
  be considerable.
 
  In addition to these considerations, comparisons between the OFS adjustment
  and the curvature adjustment of \citet{ribatet-2012a} seem natural. In our
  simulations, both adjustments performed extremely well. The curvature
  adjustment shares with OFS both the advantages and disadvantages described
  above. But in addition, the curvature adjustment, as implemented in the data
  example in \citet{ribatet-2012a}, has several additional drawbacks to
  consider. First, \citet{ribatet-2012a} require an outside method to estimate
  of $\btheta$ and $\bOmega$, whereas the OFS adjustment uses the MCMC sample to
  estimate $\btheta$ and $\bOmega$. Using the un-adjusted quasi-posterior sample
  to estimate $\btheta$ and $\bOmega$ as we do here takes advantage of borrowing
  strength, leveraging prior information, etc. that simply maximizing
  $\ell_M(\btheta)$ cannot. We note, however, that this drawback in the
  curvature adjustment can easily be avoided. One could easily apply the
  strategy that we suggest in Section \ref{sec:ofs-gibbs}, running the sampler
  first without adjustment to estimate $\btheta$ and $\bOmega$ in a
  Bayesian-like way, and then using these estimates to implement the
  curvature-adjusted sampler. In the Metropolis context, this strategy requires
  twice the computational effort of OFS; since OFS is applied to the sample post
  hoc, there is no need to run the sampler a second time. In the Gibbs sampler
  setting, however, the computational burden is identical.
  
  More obvious is the enormous computational cost imposed by estimating the
  conditional adjustment matrix in the adaptive version of the curvature
  adjusted Gibbs sampler favored by \citet{ribatet-2012a}. This simply would not
  have been feasible, for example, in the eBird example of Section
  \ref{sec:bird-counts}.  We note that this complication can probably be
  avoided by using the marginal, rather than the conditional, version of
  $\hat{\bOmega}$. In fact, since their simulated comparisons between the
  marginal and conditional forms of $\hat{\bOmega}$ performed so similarly, we
  are confused as to why \citet{ribatet-2012a} use the much more computationally
  expensive conditional version in their data example. In the end, conditional on
  implementation details, the OFS and curvature adjustments are quite similar
  both in performance and in spirit.

\section*{Acknowledgements}

  This work was funded in part by NSF grants DMS-0914906, ITS-0612031,
  ESI-0087760, and ONR grant N00244-11-1-009. The author wishes to thank Cari
  Kaufman, Richard Smith, Dan Cooley, and Alan Gelfand for much helpful
  discussion. Patient guidance on the eBird analysis was provided by Wesley
  Hochachka, Steve Kelling, and especially Daniel Fink at the Cornell Lab of
  Ornithology. Implementation of the GEVP was aided by generous
  assistance from Mathieu Ribatet.

\bibliographystyle{plainnat}
\bibliography{ofs}

\end{document}